\documentclass[letterpaper,twocolumn,pra,
aps,showpacs,groupedaddress,floatfix]{revtex4-2}

\usepackage[english]{babel}
\usepackage[utf8]{inputenc}
\usepackage[dvipsnames, hyperref]{xcolor}
\usepackage[normalem]{ulem}
\usepackage{blindtext}

\usepackage{multirow}
\usepackage{mathtools}
\usepackage{amssymb}
\usepackage{amsbsy}
\usepackage{amsmath}
\usepackage{amsfonts}
\usepackage{stmaryrd}
\usepackage{array}
\usepackage{bm}
\usepackage[arrowdel]{physics}
\usepackage{units}

\usepackage{graphicx}
\usepackage{epsfig}
\usepackage[above, below]{placeins}

\usepackage{natbib}
\usepackage{hyperref}
\hypersetup{colorlinks,linkcolor=ForestGreen,citecolor=ForestGreen,urlcolor=ForestGreen}

\addto\extrasenglish{

}

\makeatletter

\begin{document}
	\title{Quantum scars and bulk coherence in a symmetry-protected topological
		phase}
	
	\author{Jared Jeyaretnam}
	\affiliation{Department of Physics and Astronomy, University College London,
		Gower Street, London WC1E 6BT, UK}
	
	\author{Jonas Richter}
	\affiliation{Department of Physics and Astronomy, University College London,
		Gower Street, London WC1E 6BT, UK}
	
	\author{Arijeet Pal}
	\affiliation{Department of Physics and Astronomy, University College London,
		Gower Street, London WC1E 6BT, UK}
	%--------------------------------------------------------------------------------------------------
	\begin{abstract}
		Formation of quantum scars in many-body systems provides a novel mechanism for enhancing coherence of weakly entangled states.
		At the same time, coherence of edge modes in certain symmetry protected topological (SPT) phases can persist away from the ground state. 
		In this work we show the existence of many-body scars and their implications on bulk coherence in such an SPT phase.
		To this end, we study the eigenstate properties and the dynamics of an interacting spin-$1/2$ chain with three-site ``cluster'' terms hosting a $\mathbb{Z}_2 \times \mathbb{Z}_2$ SPT phase.
		Focusing on the weakly interacting regime, we find that eigenstates with volume-law entanglement coexist with area-law entangled eigenstates throughout the spectrum.
		We show that a subset of the latter can be constructed by virtue 
		of repeated cluster excitations on the even or odd sublattice of the 
		chain, resulting in an equidistant ``tower'' of states, analogous to the        
		phenomenology of quantum many-body scars.
		We further demonstrate that these scarred eigenstates support 
		nonthermal expectation values of local cluster operators in the bulk and exhibit 
		signatures of topological order even at finite energy densities.
		Studying the dynamics for out-of-equilibrium states drawn from the 
		noninteracting ``cluster basis'', we unveil that nonthermalizing bulk dynamics 
		can be observed on long time scales if clusters on odd and even sites are 
		energetically detuned.
		In this case, cluster excitations remain essentially confined to one of the two sublattices such that inhomogeneous cluster configurations cannot equilibrate and thermalization of the full system is impeded.
		Our work sheds light on the role of quantum many-body scars in preserving SPT order at finite temperature and the possibility of coherent bulk dynamics in models with SPT order beyond the existence of long-lived edge modes.
		
	\end{abstract}
	
	\maketitle
	%---------------------------------------------------------------------------------------------------
	
	\section{Introduction}\label{sec:introduction}
	
	The out-of-equilibrium dynamics of many-body quantum systems has been of great interest for a number of years now \cite{Polkovnikov_2011, Eisert_2015}.
	While a generic interacting quantum system is expected to thermalize and lose all local memory of its initial conditions \cite{D_Alessio_2016, Borgonovi_2016}, it is an active frontier of modern theoretical and experimental physics to identify effective mechanisms which can impede this thermal fate and allow for coherent quantum dynamics on long time scales.
	In fact, various examples are known by now where thermalization can be avoided.
	One such example is integrability, meaning that a system has an extensive number of conserved quantities which prevent equilibration to standard ensembles of statistical mechanics \cite{Vidmar_2016, Essler_2016}.
	While integrable models represent isolated points in parameter space, the concept of many-body localization (MBL) in strongly disordered systems provides a means to break ergodicity for non-finetuned models as well \cite{Nandkishore_2015, Abanin_2019}.
	
	While the question of thermalization is usually concerned with states at finite energy densities or even at \textit{infinite} temperature in the middle of the spectrum, symmetry-protected topological (SPT) phases at {\it zero} temperature are well known to host robust edge modes which cease to decay due to an energy gap in the bulk \cite{Gu_2009, Chen_2011, Fidkowski_2011, Levin_2012, Pollmann_2012, Senthil_2015, Parker_2019}, and such phases have been detected experimentally \cite{DeLeseleuc2019, Sompet2021}.
	At finite temperatures, on the contrary, these edge modes are expected to quickly decohere due to interactions with thermal excitations.
	Remarkably, however, strong disorder and the onset of MBL have been shown to provide a means to stabilize long-lived edge degrees of freedom also at nonzero temperatures \cite{Bahri2013}.
	In this case, the lifetime of the edge mode increases exponentially with the size of the system such that in the thermodynamic limit, the edge mode stays coherent on indefinite time scales.
	Similarly, it was recently demonstrated in Ref.~\cite{Kemp2019} that such prethermal edge qubits can also persist in certain parameter regimes of disorder-free SPT models thanks to the presence of (almost) strong zero modes \cite{Fendley_2016, Kemp_2017, Else_2017, Vasiloiu_2018, Fendley_2012}.
	Specifically, Ref.~\cite{Kemp2019} introduced a ``dimerization'' parameter which causes a decoupling of bulk and boundary, leading to exponentially long coherence times of the edge mode.
	In the present paper, we extend the investigations of Ref.~\cite{Kemp2019} and show that the dimerized model can additionally host anomalously long-lived dynamics in the bulk of the system.
	
	Thermalization in isolated quantum system is often understood in terms of the \textit{eigenstate thermalization hypothesis} (ETH) \cite{D_Alessio_2016, Srednicki1994, Rigol_2008, Deutsch_2018}.
	In essence, ETH asserts that expectation values of physical operators evaluated with respect to individual eigenstates form a smooth function of energy and agree with the corresponding microcanonical ensemble average.
	While the ETH is clearly violated in integrable and many-body localized systems \cite{Nandkishore_2015}, there has been substantial numerical evidence that the ETH is fulfilled in a multitude of generic (nonintegrable) models \cite{D_Alessio_2016, Beugeling_2014, Steinigeweg_2013, Hyungwon2014, Mondaini_2016, Jansen_2019, Brenes_2020, Richter_2020, Torres2014}.
	More recently, intermediate cases which fall outside the paradigms of ``fully ETH'' or ``fully MBL'' have received an upsurge of interest.
	In particular, fascinating experiments on Rydberg atoms led to the discovery of so-called quantum many-body scars \cite{Bernien2017a, Turner2018a}.
	These scars are rare ETH-violating states which are embedded in an otherwise thermal spectrum and typically make up a vanishing proportion in the thermodynamic limit \cite{Shiraishi2017}.
	
	The presence of such scar states means that a system will thermalize for most initial conditions, but when initialized in certain specific states (which are often experimentally accessible), atypical dynamics are observed \cite{Turner2018a}.
	Subsequent work has substantiated the existence of quantum scars in a variety of models~\cite{Iadecola2019, Schecter2019, Chattopadhyay2019, Moudgalya2017, Moudgalya2018, Moudgalya2020, Scherg2020, Kuno2020, Desaules2021}.
	In such cases, scars often appear in the form of a ``tower of states'', i.e., a set of eigenstates with almost equidistant energy spacing forming a nonthermalizing subspace which can be constructed by applying certain raising-type operators~\cite{Mark2020, ODea2020}.
	
	Furthermore, there are examples of (weak) ergodicity breaking, where kinetic constraints, additional conservation laws, or high levels of frustration can cause the emergence of slow dynamics or even a fragmentation of the Hilbert space into disconnected subspaces~\cite{Pancotti_2020, Lee_2020, Lee_2020_2, McClarty_2020, Khemani_2020, Sala_2020, Magnifico2020}.
	
	While quantum many-body scars exhibiting topological order have been constructed for certain models~\cite{Ok2019, Wildeboer2020, Srivatsa2020}, their impact on the nonequilibrium dynamics in models hosting SPT phases is far less explored.
	In the present paper, we provide evidence that scarred eigenstates 
	indeed appear naturally in certain SPT models and that these nonthermal states 
	provide a means to preserve topological order even at finite energy densities, 
	similar to eigenstates of many-body localized systems~\cite{Huse2013, 
		Chandran_2014}.
	Moreover, we demonstrate the occurrence of nonthermalizing dynamics for certain operators and initial states, not only at the edges but even in the \textit{bulk} of the system.
	To this end, we consider an interacting spin-$1/2$ chain with three-site ``cluster'' terms hosting a $\mathbb{Z}_2 \times \mathbb{Z}_2$ SPT phase~\cite{Raussendorf_2001, Keating_2004, Kopp_2005, Son2011, Verresen2017} (often called the ZXZ model), which has also recently gathered interest in the context of thermalization~\cite{Bahri2013, Kemp2019}.
	\begin{figure}[tb]
		\centering
		\includegraphics[width=1\columnwidth]{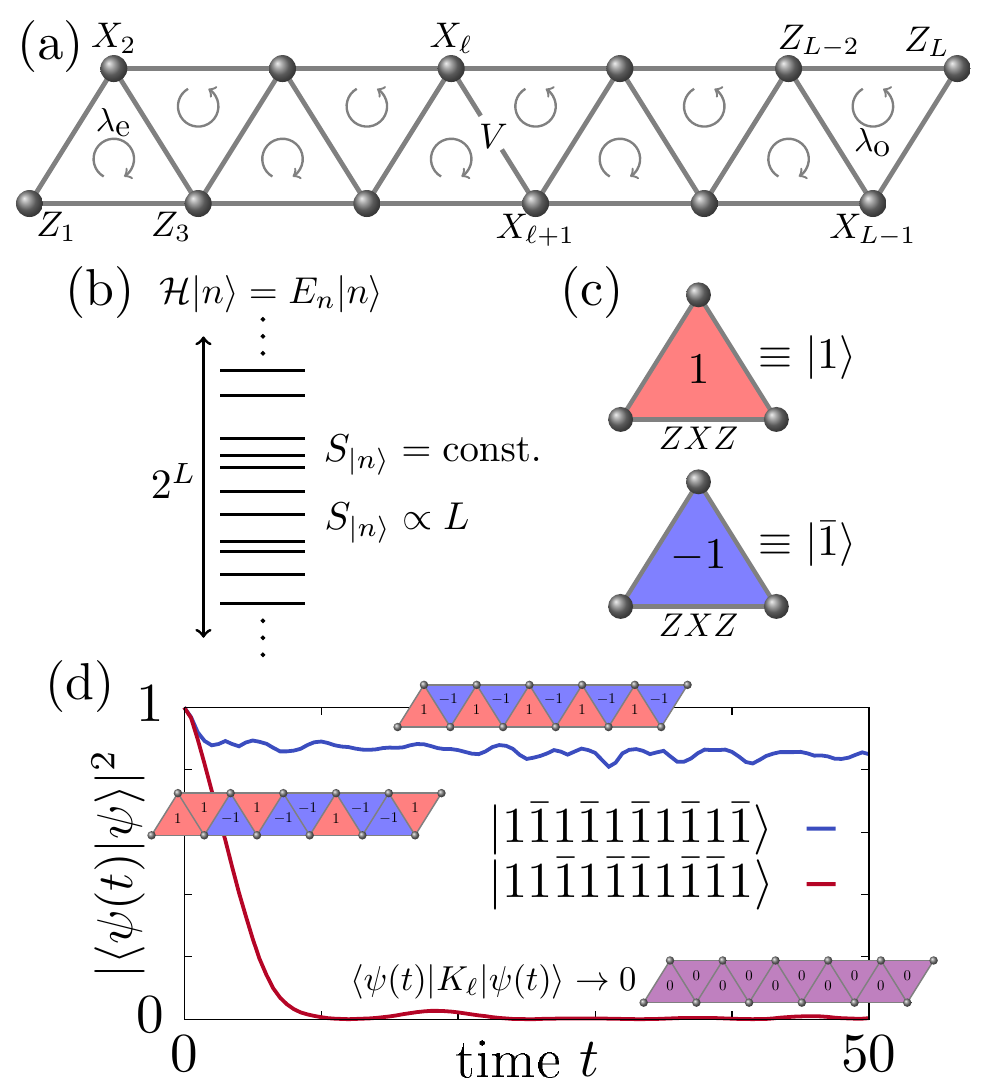}
		\caption{
			\textbf{(a)}~The Hamiltonian \eqref{eq:Ham} can
			be interpreted as a triangular ladder with the two legs
			comprising the even and
			odd lattice sites respectively.
			The integrability-breaking perturbation of strength $V$ acts on the rungs.
			Note that the global field $\Gamma\sum_\ell X_\ell$ is not shown here, and that $X_\ell$, $Z_\ell$ denote the usual Pauli operators $\sigma^{x, z}_\ell$ at site $\ell$.
			\textbf{(b)}~For the weakly interacting limit,
			we show that there is a coexistence throughout the
			spectrum between eigenstates
			with a volume-law scaling of the entanglement entropy
			and nonthermal
			eigenstates which follow an area law.
			\textbf{(c)}~ We show that a full basis of the
			Hilbert space can be constructed from states with
			definite expectation values
			$\pm 1$ of $K_\ell$ for $2\leq \ell \leq L-1$, see
			Appendix~\ref{app::cluster_basis}.
			These ``cluster-basis'' states are eigenstates of the clean ZXZ model ($\Gamma = V = 0$).
			\textbf{(d)}~Loschmidt echo $|\langle
			\psi(t)|\psi\rangle|^2$ for two different
			exemplary initial states from the
			cluster basis showing drastically different dynamics.
			The parameters in (d) are chosen as $L = 12$, $\Gamma = 0.1$, $V = 0.05$, $\lambda_\text{e} = 0.6$, and $\lambda_\text{o} = 1$.
			Consistent with Eq.~\eqref{eq:Ham}, we denote the dimerization parameter as $\lambda_\text{e} = \lambda$ and drop $\lambda_\text{o} = 1$ in the following.
		}
		\label{fig:chain_diagram}
	\end{figure}

	The ZXZ chain [as defined below in Eq.~\eqref{eq:Ham}] can also be understood as a triangular ladder with the two legs comprising the even and odd lattice sites of the chain respectively, as well as an integrability-breaking interaction acting on the rungs; see Fig.~\ref{fig:chain_diagram}(a).
	Focusing on the weakly interacting regime, we study the eigenstate properties of this model and unveil that volume-law entangled eigenstates coexist with area-law entangled eigenstates throughout the spectrum; see Fig.~\ref{fig:chain_diagram}(b).
	While we refrain from classifying all the putative area-law entangled states, we show that a subset of them can be constructed in terms of a ``tower'' of states, similar to the phenomenology of quantum many-body scars.
	Here, this tower is constructed by virtue of repeated cluster excitations on one of the two sublattices, with the cluster operators on the other sublattice all remaining in their ground state.
	Motivated by this construction we study the real-time dynamics of out-of-equilibrium states with definite expectation values $\pm 1$ of local clusters, i.e., eigenstates of the ZXZ model for vanishing interactions; see Fig.~\ref{fig:chain_diagram}(c) and
	Appendix~\ref{app::cluster_basis}.
	If clusters on odd and even sites are energetically detuned due to dimerization \cite{Kemp2019}, we demonstrate that some of these states exhibit atypical nonthermalizing dynamics on long time scales.
	This is exemplified in Fig.~\ref{fig:chain_diagram}(d) by the Loschmidt echo ${\cal L}(t)$,
	\begin{equation}\label{eq:loschmidt}
		{\cal L}(t) = |\langle \psi(t)|\psi\rangle|^2\ ,\quad \ket{\psi(t)} = e^{-i{\cal H}t}\ket{\psi}\ .
	\end{equation}
	Specifically, ${\cal L}(t)$ is found to decay very slowly for an initial state $\ket{\psi}$ where all clusters of the even sublattice are excited, while the clusters of the odd sublattice are in the ground state [blue curve in Fig.~\ref{fig:chain_diagram}(d)].
	We argue that the slow dynamics can be understood as an effective restriction of the cluster excitations to one of the two sublattices, such that inhomogeneous cluster configurations cannot equilibrate and thermalization of the full system is impeded.
	As a consequence, such initial states, where the clusters of one sublattice are all either in the ground state or all in the excited state are found to be particularly stable.
	In contrast, other initial states with a finite number of excitations on both sublattices can thermalize quickly [red curve in Fig.~\ref{fig:chain_diagram}(d)].
	
	This paper is structured as follows.
	First, in Sec.~\ref{sec:Model}, we provide an introduction to the
	model studied in this work.
	In Sec.~\ref{sec:scars}, we then characterize the properties of the
	scarred eigestates and show that they support nonthermal expectation
	values of cluster operators in the bulk and exhibit signatures of
	topological order.
	In Sec.~\ref{sec:Dynamics}, we turn to the dynamical aspects of
	the ZXZ model where we particularly focus on quenches starting from
	eigenstates of the noninteracting model.
	We close with a discussion of our results in Sec.~\ref{sec:Discussion},
	where we also mention some directions of future research.
	
	\section{The model}\label{sec:Model}
	
	We study a generalized version of the ZXZ or ``cluster'' model which, in addition to the bare cluster terms, also includes an integrability-breaking perturbation and a global transverse field.
	The Hamiltonian for a chain with open boundary conditions (OBC) and $L$ lattice sites is given by \cite{Kemp2019} [see also Fig.~\ref{fig:chain_diagram}(a)],
	\begin{equation}
		\label{eq:Ham} {\cal H} = \hspace{-0.2cm}\sum_{\ell = 1}^{L/2-1} \left(\lambda K_{2\ell} + K_{2\ell+1}\right) + \Gamma \sum_{\ell = 1}^L X_\ell + V\sum_{\ell=1}^{L-1}X_\ell X_{\ell+1} ,
	\end{equation}
	where the cluster operators $K_\ell$ are defined as,
	\begin{equation}
		\label{eq:Cluster} K_\ell = Z_{\ell-1}X_\ell Z_{\ell+1}\ ,
	\end{equation}
	and in this work $X_\ell$, $Y_\ell$, $Z_\ell$ denote the usual Pauli operators $\sigma^{x, y, z}_\ell$ at site $\ell$.
	$\mathcal{H}$
	obeys a $\mathbb{Z}_2 \times \mathbb{Z}_2$ symmetry \cite{Son2011},
	consisting
	of spin inversion symmetry on odd and even sites respectively.
	When $\lambda = 1$, the model also has an additional $\mathbb{Z}_2$ parity symmetry corresponding to inverting the chain about the central site.
	Note that, throughout this paper, it is understood that while ${\cal H}$ has $L$ lattice sites with subscripts $1\leq \ell \leq L$, due to OBC there are only $L-2$ cluster terms $K_\ell$ labeled by $2\leq \ell \leq L-1$.
	
	The cluster operators $K_\ell$ in Eq.~\eqref{eq:Cluster} are mutually commuting and have eigenvalues $\pm1$; see Fig.~\ref{fig:chain_diagram}(c).
	Therefore, they define a complete basis of states within each symmetry sector (see Appendix~\ref{app::cluster_basis}), labeled by the eigenvalues of the cluster operators on each site.
	These states are known in the quantum information literature as ``cluster states'' \cite{Raussendorf_2001}.
	When $\Gamma = V = 0$ (i.e.\ the clean ZXZ model), these states are exact eigenstates of ${\cal H}$, and the ground state $|\text{gs}\rangle$ in each sector is the state with,
	\begin{equation}
		\label{eq:ground_state} \matrixel{\text{gs}}{K_{\ell}}{\text{gs}} = -1\ ,\ \quad 2\leq \ell \leq L-1\ .
	\end{equation}
	Below in Sec.~\ref{sec:Dynamics}, we will study the dynamics of such cluster-basis states for quenches in the interacting model with $\Gamma\neq 0$, $V \neq 0$.
	Note that in Fig.~\ref{fig:chain_diagram} and also further below we label the cluster-basis states $\ket{\psi}$ by their expectation values of the $K_\ell$, e.g., $\ket{\psi} = |1\bar{1}1\cdots\rangle$.
	
	While the spectrum of ${\cal H}$ is trivial in the case of $\Gamma = V = 0$, ${\cal H}$ in fact still remains integrable for $\Gamma \neq 0$.
	In this case, the model can be recast in terms of fermionic operators by means of a Jordan-Wigner transform~\cite{Son2011, Huse2013, Verresen2017, Kemp2019}, where the three-spin cluster operators become next-nearest-neighbor interactions between fermions.
	This means that the even sites and odd sites become entirely disconnected, and the model can be separated into two copies of the transverse-field Ising model (TFIM),
	${\cal H}(V = 0) \sim {\cal H}^{\mathrm{e}}_{\text{TFIM}}(\lambda, \Gamma) + {\cal H}^{\mathrm{o}}_{\text{TFIM}}(\lambda_\text{o} =1, \Gamma)$, which is well known to be integrable~\cite{Pfeuty_1970},
	\begin{equation} \label{eq:single_chain}
		\mathcal{H}^{\hspace{0.08em}\mathrm{e/o}}_{\text{TFIM}}(\lambda, \Gamma) =
		\lambda \hspace{-0.6em} \sum_{\ell\ \in\ \mathrm{e/o}} \hspace{-0.7em}  Z_\ell Z_{\ell+2}
		+ \Gamma \hspace{-0.6em} \sum_{\ell\ \in\ \mathrm{e/o}} \hspace{-0.7em} X_\ell\ ,
	\end{equation}
	where the sums in Eq.~\eqref{eq:single_chain} should
	be understood as either running over the even (e) or odd (o) lattice
	sites.
	
	However, the introduction of a nonzero $V$ adds
	couplings between these two
	chains, breaking their integrability.
	The model in Eq.~\eqref{eq:Ham} can hence be viewed as a ladder, with one leg given by the even sites and the other by the odd sites, as shown in Fig.~\ref{fig:chain_diagram}.
	In this work, we focus on the weakly interacting regime and choose $\Gamma = 0.1$ and $V = 0.05$, similar to Refs.~\cite{Bahri2013, Kemp2019}.
	While ${\cal H}$ becomes nonintegrable in this case, we find that, at least for the system sizes $L$ which are numerically available, ${\cal H}$ is not strongly thermalizing but rather exhibits an intermediate behavior between integrability and full quantum chaos.
	This fact reflects itself for instance in comparatively broad distributions of eigenstate entanglement entropies (Sec.~\ref{sec:entropy}) and
	eigenstate expectation values of cluster operators $K_\ell$
	(Sec.~\ref{sec:ClusterExpV}), as well as in the distribution
	of adjacent level spacings (see Appendix~\ref{App::Chaos}).
	
	The ZXZ model \eqref{eq:Ham} exhibits symmetry-protected topological order, protected by the $\mathbb{Z}_2 \times \mathbb{Z}_2$ spin-flip symmetry \cite{Son2011}.
	As a result, the model hosts robust boundary degrees of freedom at zero temperature.
	While such edge modes typically decohere at finite temperature, due to interaction with thermal excitations in the bulk, Ref.~\cite{Bahri2013} showed that they can be stabilized by means of strong disorder and the onset of MBL.
	In this case, the life time of the zero modes increases exponentially with the size of the system such that in the thermodynamic limit, $L \to \infty$, the boundary qubit remains coherent on indefinite time scales.
	Moreover, even without disorder, it has been demonstrated in Ref.~\cite{Kemp2019} that such prethermal boundary modes can survive at infinite temperature in certain parameter regimes if $\lambda \neq 1$ in Eq.~\eqref{eq:Ham}.
	(Note that $\lambda \neq 1$
	breaks the
	$\mathbb{Z}_2$ ``swap'' symmetry between
	chains while still preserving the SPT-protecting $\mathbb{Z}_2 \times
	\mathbb{Z}_2$ spin-flip symmetry.) This choice leads to a
	``dimerization'' of the chain such that each leg of the ladder has a
	different
	cost for excitations.
	Motivated by this work, we here set $\lambda = 0.6$ (as in Ref.~\cite{Kemp2019}) and scrutinize the eigenstate properties and the dynamics of ${\cal H}$.
	In particular, we provide evidence that the spectrum of $\mathcal{H}$ 
	hosts scarred sub-volume law entangled eigenstates which exhibit signatures of 
	SPT order at finite energy densities.
	Moreover, we demonstrate that in the same parameter regime where Ref.~\cite{Kemp2019} reported the existence of long-lived edge modes, the bulk dynamics becomes anomalous as well, at least when considering appropriate operators and initial states.
	Specifically, for cluster-basis states $\ket{\psi}$ with a particular initial configuration $\mel{\psi}{K_\ell}{\psi}$ of clusters, we observe long coherence times as cluster excitations essentially remain confined to one of the two sublattices.
	
	\section{Nonthermal eigenstates in the ZXZ model}\label{sec:scars}
	Quantum many-body scars are states which violate the ETH in an otherwise chaotic system, despite having finite energy density.
	This stands in contrast with integrable and many-body localized systems, in which every state violates the ETH, and fully chaotic systems, in which every state with finite energy density is thermal.
	In this section, starting from low-energy excitations in the regime of perturbatively small $\Gamma$ and zero $V$, we will show that a tower of nonthermal eigenstates exists in the ZXZ model, which is well approximated by a generalization of the low-energy excitations.
	We provide a detailed characterization of these eigenstates and demonstrate that they feature low entanglement entropies and atypical expectation values of local cluster operators.
	Furthermore, we will show that these states may preserve a four-fold degenerate entanglement spectrum, a key signature of $\mathbb{Z}_2 \cross \mathbb{Z}_2$ SPT order.
	\subsection{Low-energy excitations}
	Because the non-interacting ($V = 0$) Hamiltonian can be separated into two copies of the transverse-field Ising model, corresponding to the odd and even sublattices respectively, the excitations of the model should be equivalent in both cases.
	In the regime we investigate, the ZXZ model maps onto the ferromagnetic phase of the TFIM -- that is, where $\lambda \gg \Gamma$ in Eq.~\eqref{eq:single_chain}.
	In this phase of the TFIM, excitations are given by domain walls between two regions of aligned spins, while the $X_\ell$ term causes these walls to hop, resulting in delocalized excitations~\cite{Huse2013}. 
	The domain walls in the TFIM map onto cluster excitations in the ZXZ model~\cite{Son2011,Huse2013, Kemp2019} and by expressing a bulk $X_\ell$ operator as,
	\begin{equation}
		\left(K^+_{\ell-1} + K^-_{\ell-1}\right)K_\ell\left(K^+_{\ell+1} + K^-_{\ell+1}\right) = X_\ell\ ,
	\end{equation}
	where $K_\ell^\pm$ are cluster raising and lowering operators,
	\begin{equation}\label{eq:raising_operator}
		K^\pm_\ell =
		\frac{1}{2} \left(Z_\ell \mp i Z_{\ell-1} Y_\ell Z_{\ell+1}\right)\ ,
	\end{equation}
	it is clear that the $X_\ell$ term flips two next-nearest cluster operators.
	This will cause an isolated excitation to hop two sites at a time, at zero energy cost.
	The term can also create or destroy two such excitations, however this changes the energy of the state.
	Considering a perturbatively small $\Gamma$, the zero-energy parts of this term mix the cluster states under degenerate perturbation theory, with a first order effect in the energy, while the other parts act only to first order on the state and second order on the energy.
	Therefore, in the regime of very small $\Gamma$, we should expect that the low-lying eigenstates in the ZXZ model with periodic boundary conditions resemble delocalized cluster excitations.
	A full demonstration of this can be found in Appendix~\ref{app::excitations}, but the relevant result is that there exist low-lying states $\ket{k}$ with energy $\epsilon_k = 2 - 2\Gamma\cos(k)$,
	\begin{equation}
		\label{eq:kstate}
		\ket{k} = \sqrt{\frac{2}{L}} \,
		\sum_{\ell = 0}^{L/2 - 1}\! e^{+ik\ell} \ket{2\ell +
			1}\ ,
	\end{equation}
	with $k = 2\pi \eta/(L/2)$, $0 \leq \eta < L/2$.
	$\ket{2\ell + 1} = K_{2\ell + 1}^+ \ket{\text{gs}}$ describes a localized cluster excitation at site $2\ell + 1$, such that $\ket{k}$ describes a delocalized cluster excitation with momentum $k$ on the odd sublattice.
	While the excitations $\ket{k}$ in Eq.~\eqref{eq:kstate} apply to periodic boundary conditions, perturbatively small $\Gamma$, and $V = 0$, we find that, with some modifications, these states also convincingly approximate
	low-lying excitations in the model with OBC and nonzero $V$.
	More surprisingly, we find that states with multiple such excitations provide good approximations even for eigenstates closer to the center of the spectrum.
	\subsection{Approximate tower of states}\label{sec:tower}
	Working with OBC, the sum over $\ell$ in Eq.~\eqref{eq:kstate} now starts at $\ell = 1$, as the first odd cluster site is $K_{2\ell + 1} = K_3$, cf.\ Eq.~\eqref{eq:Ham}. Moreover, the allowed values of $k$ are $k = {2\pi \eta}/({L/2 - 1})$,
	$0 \leq \eta < L/2 - 1$, and the normalization changes.
	With these changes, we find that the states
	$\left(\ket{+k} \pm \ket{-k}\right)/\sqrt{2}$ each have a high overlap with a particular eigenstate of the model, for every allowed value of $k$, even with $V > 0$.
	
\begin{figure}[tb]
		\centering
		\includegraphics[width=0.85\columnwidth]{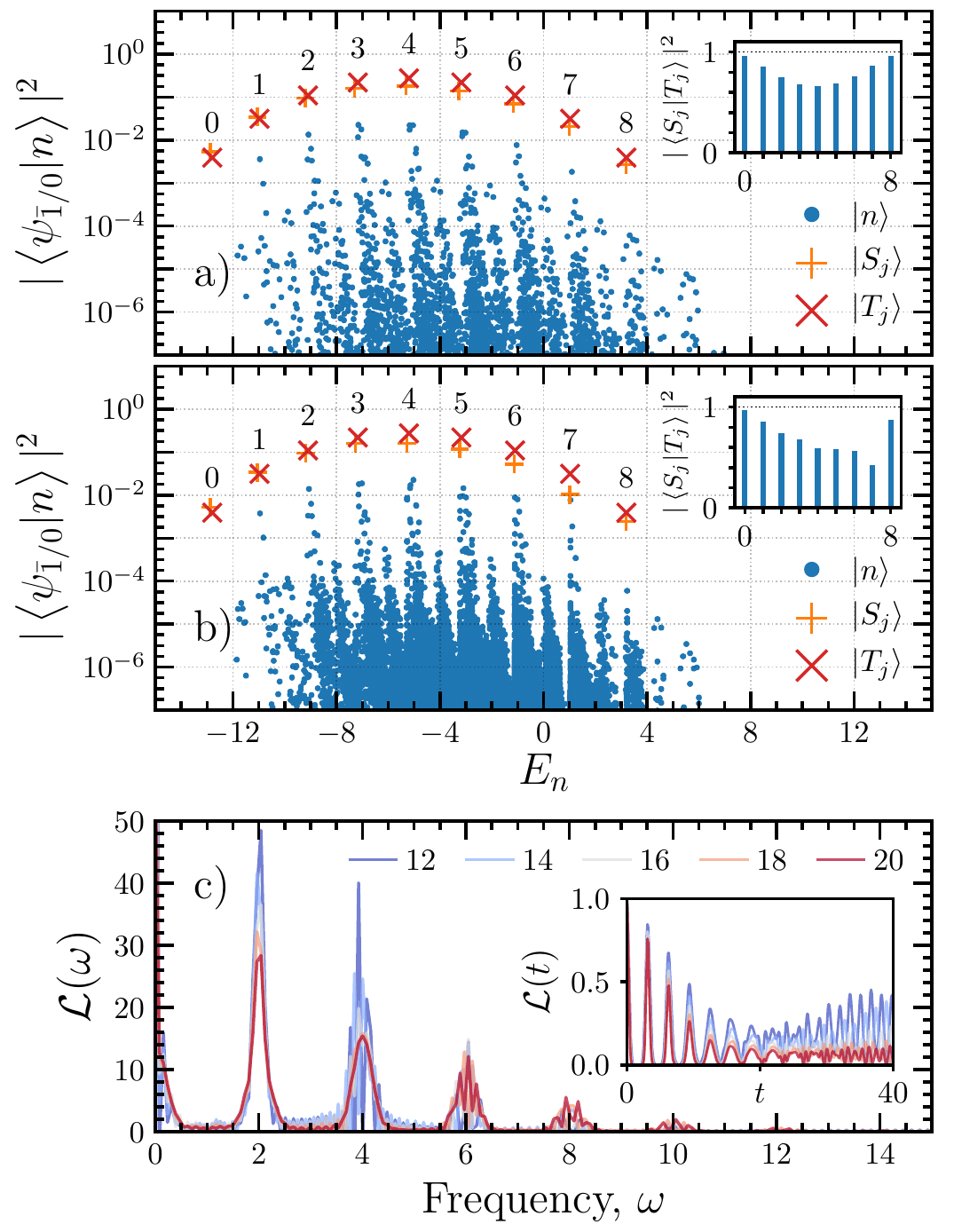}
		\caption{
			\textbf{[(a),~(b)]}
			Squared overlaps of the eigenstates of $\mathcal{H}$ with the state $\ket{\psi_{\bar{1}/0}}$, plotted against energy, for $L = 18$, in the subspace with positive spin-flip symmetry on both sublattices.
			Panel (a) shows the results for the non-interacting case ($V =0$) and (b) for the interacting case ($V = 0.05$).
			The highlighted eigenstates $\ket{S_j}$ exhibit a significant and atypically high overlap with $\ket{\psi_{\bar{1}/0}}$, and can be well approximated by the tower of states $\ket{T_j}$ [see insets to panels (a) and (b)].
			\textbf{(c)} Fourier-transformed Loschmidt echo \eqref{eq:loschmidt} of the state $\ket{\psi_{\bar{1}/0}}$ for system sizes from $L=12$ up to $L=20$, with $V = 0.05$.
			The inset shows the original time-domain data. 
			[$\lambda = 0.6$, $\Gamma = 0.1$ in all cases.]
		}
		\label{fig:tower}
	\end{figure}

	To generalize this result to multiple excitations, we will introduce an operator which creates a single delocalized excitation at a time.
	To start with, consider the cluster raising (lowering) operator \eqref{eq:raising_operator}, which creates (destroys) an excitation at a given site.
	Using this and focusing on excitations at zero momentum ($k = 0$), we then construct an operator acting upon the entire chain by,
	\begin{equation}
		\label{eq:tower_operator} \mathcal{O} = 
		\! \sum_{\ell =1}^{L/2 - 1}\! \! K^+_{2\ell + 1}\ .
	\end{equation}
	We then apply the operator repeatedly to the ground state $\ket{\text{gs}}$~\eqref{eq:ground_state} of the clean ZXZ model to produce a tower of states~\cite{Moudgalya2017, Iadecola2019, Chattopadhyay2019, Schecter2019, Mark2020, ODea2020, Moudgalya2020}.
	That is, we generate a set of states $\ket{T_j}$ given by,
	\begin{equation}
		\ket{T_j} = \frac{\mathcal{O}^j \ket{\text{gs}}}{\left|\mathcal{O}^j \ket{\text{gs}}\right|}.
	\end{equation}
	Because the operator \eqref{eq:tower_operator} produces a single delocalized excitation, each set of states has an energy spacing of approximately $2$ -- this spacing is exact in the $\Gamma = V = 0$ model.
	In the insets of Fig.~\ref{fig:tower}(a)~and~(b), we show that the states $\ket{T_j}$ have large overlaps with certain eigenstates of ${\cal H}$ in both the non-interacting ($V = 0$) and the interacting ($V = 0.05$) case.
	
	In the following, we label those eigenstates $\ket{n}$ which maximize the overlap $\left|\braket{n}{T_j}\right|^2$ for a given tower index $j$ by $\ket{S_j}$.
	In particular, the comparatively simple structure of the 
	$\ket{T_j}$ already indicates that the eigenstates $\ket{S_j}$ might exhibit 
	atypical properties, which we will analyze in more detail below.
	Note, however, that $\ket{T_7}$ in fact has large and comparable overlaps with a pair of adjacent eigenstates, but it is only the one with the larger overlap that we label $\ket{S_7}$.
	This feature of $\ket{T_7}$ may indicate that the corresponding scar state loses stability for increasing $L$, as this was not observed for any of the $\ket{T_j}$ in the $L=14$ case and is only weakly apparent for $L=16$ for certain $j$.
	
	If an initial state $\ket{\psi}$ can be found whose spectral 
	decomposition is dominated by the scarred eigenstates $\ket{S_j}$, such a 
	$\ket{\psi}$ should yield periodic oscillations in time.
	As shown in Fig.~\ref{fig:tower}(a)~and~(b), such a state indeed exists and can be constructed by applying cluster-lowering operators on the even cluster sites to a spin-basis product state (for details, see Appendix~\ref{app::cluster_basis}).
	The resulting state $\ket{\psi_{\bar{1}/0}}$ is a superposition of cluster basis states and has expectation values,
	\begin{equation}
		\matrixel{\psi_{\bar{1}/0}}{K_{2\ell}}{\psi_{\bar{1}/0}} = -1,\ \matrixel{\psi_{\bar{1}/0}}{K_{2\ell + 1}}{\psi_{\bar{1}/0}} = 0\ ,
	\end{equation}
	although it is important to note that these expectation values 
	alone do not define this state uniquely (see Appendix~\ref{app::cluster_basis}).
	In Fig.~\ref{fig:tower}~(c), we show the Fourier-transformed 
	Loschmidt echo ${\cal L}(\omega) = {\cal F}[{\cal L}(t)]$ of the state 
	$\ket{\psi_{\bar{1}/0}}$. As expected from the high overlap with the almost 
	equidistant states $\ket{S_j}$, ${\cal L}(\omega)$ exhibits peaks at 
	frequencies $\omega$ which are multiples of $2$. For increasing 
	system size $L$, the spectral contributions of peaks at higher $\omega$ become 
	slightly more pronounced. Correspondingly, we find that the revivals of ${\cal 
		L}(t)$ in time [see inset of Fig.~\ref{fig:tower}~(c)] become  
	less distinct for increasing $L$. 
	
	Note that in addition to the $\ket{T_j}$ discussed above, there are in fact three other towers starting from the ground state of each symmetry sector.
	These can be obtained by replacing the raising operators in Eq.~\eqref{eq:tower_operator} with lowering operators and starting from the most excited state, or by applying the operators to even sites, or by a combination of both.
	Physically, these correspond to the transformations $K_\ell \rightarrow -K_\ell$ and $K_{2\ell} \leftrightarrow K_{2\ell+1}$ respectively.
	Moreover, while the operator $\mathcal{O}$ produces a zero-momentum excitation, we find that generalizations to $k \neq 0$ as well as to half-integer values of $\eta$ seem to also yield good approximations to eigenstates of ${\cal H}$ (see Appendix~\ref{App::kdata} for details).
	Interestingly, if we start the construction of the tower from some state besides the ground state (e.g.\ with a single excitation on the   even sites), the above results do not hold, i.e., the states generated are poor matches to eigenstates of ${\cal H}$.
	
	Eventually, let us mention one technical detail.
	Specifically, the operator $\mathcal{O}$ in Eq.~\eqref{eq:tower_operator} changes the $\mathbb{Z}_2$ spin-flip symmetry on the odd sublattice such that the tower produced starting from a state in one symmetry sector will alternate between that and one other symmetry sector.
	For computational convenience, we choose to shift the entire tower into the same symmetry sector.
	This is achieved by applying the operator $Z_1$ after every application of the operator $\mathcal{O}$, as $Z_1$ anticommutes with the spin flip symmetry on odd sites, but commutes with all cluster operators.
	As the states $\ket{T_j}$ in each symmetry sector are equivalent, this should not affect the results, and we have checked this by performing the same analysis in different symmetry sectors.
	\subsection{Entanglement entropy}\label{sec:entropy}
	Given a state $\ket{\psi}$, its entanglement entropy for a bipartition into subsystems $A$ and $B$ is given by,
	\begin{equation}
		\label{eq:entropy} S_A = -\text{Tr}[\rho_A \ln \rho_A]\ ,\quad \rho_A = \text{Tr}_B\lbrace\dyad{\psi}{\psi}\rbrace\ ,
	\end{equation}
	where $\rho_A$ is the reduced density matrix for a subsystem~$A$.
	For an eigenstate of a Hamiltonian obeying the ETH, one generally expects that $S_A$ scales with the system size, and in particular at infinite temperature it approaches the Page value~\cite{Page1993}, which is the average entropy for a random pure state.
	A state with an extensive entanglement entropy is said to be obeying a volume-law.
	In contrast, the ground state of gapped systems is always area-law~\cite{Eisert2010}, even in chaotic systems.
	Surprisingly, states with sub-extensive entanglement entropies have been found even at finite energy densities in a number of otherwise chaotic models, now usually referred to as quantum many-body scars~\cite{Turner2018a}.
	\begin{figure}[tb]
		\centering
		\includegraphics[width=0.92\columnwidth]{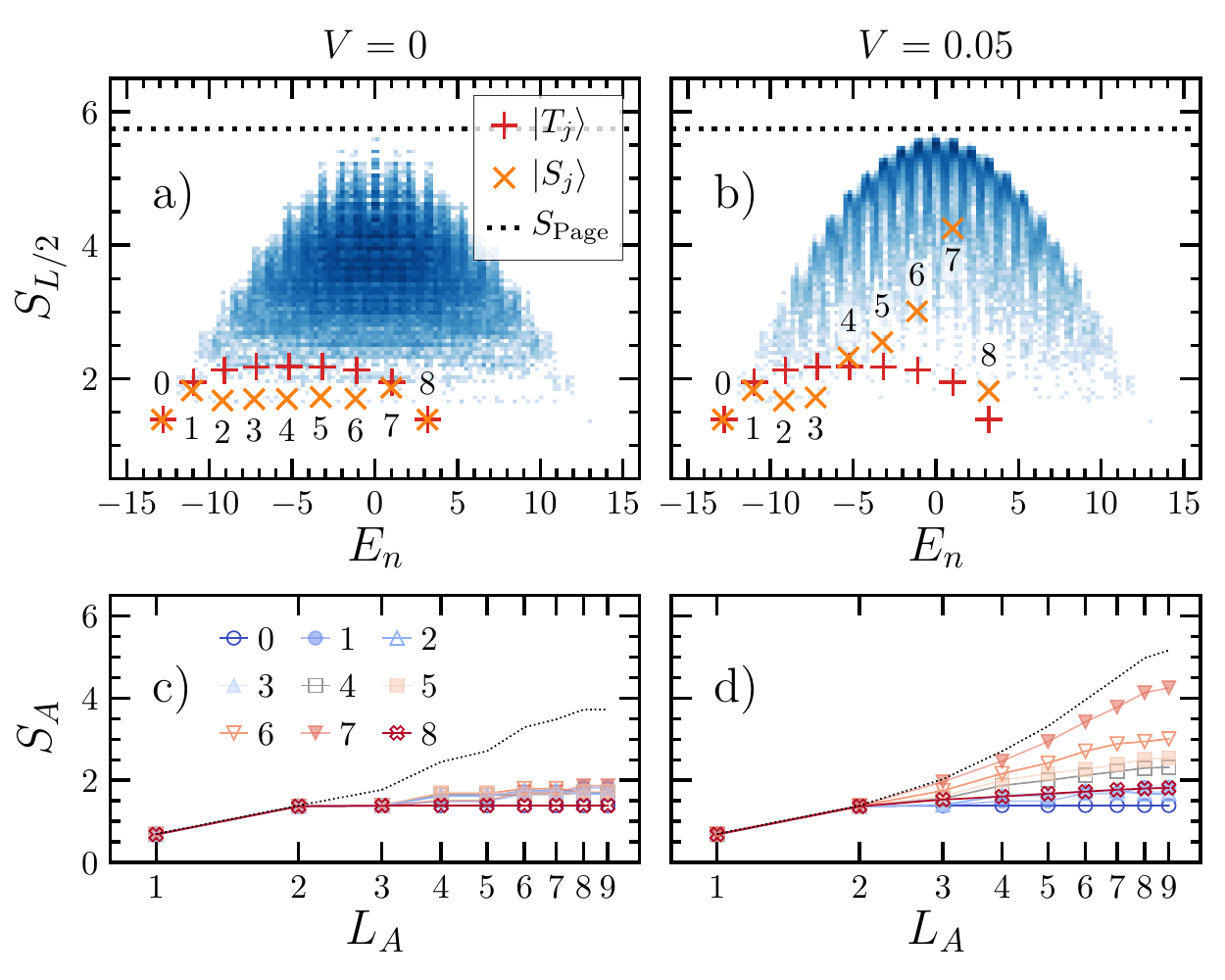}
		\caption{
			\textbf{(a)}
			Half-system bipartite entanglement entropy of the eigenstates of the non-interacting ($V = 0$) Hamiltonian, plotted against energy, for $L=18$ in the subspace with positive spin-flip symmetry on both sublattices.
			Darker colors indicate a greater number of states at that value.
			Additionally, the tower of states $\ket{T_j}$ and their corresponding eigenstates $\ket{S_j}$ are highlighted by crosses.
			\textbf{(b)} The same as (a), but for the interacting ($V = 0.05$) model.
			\textbf{(c)} The scaling of entanglement entropy with subsystem size in the non-interacting model is shown for the states $\ket{S_j}$.
			A typical thermal eigenstate (dotted black line) exhibiting a volume law is shown for comparison.
			\textbf{(d)} The same as (c), but for the interacting ($V = 0.05$) model.
			[$\lambda = 0.6$, $\Gamma = 0.1$ in all cases.]
		}
		\label{fig:entropy}
	\end{figure}
	
	Figures~\ref{fig:entropy}(a)~and~(b) show the half-system bipartite entanglement entropy for the eigenstates of the ZXZ model \eqref{eq:Ham} when $L_A = L/2$, in both the non-interacting ($V = 0$) and the interacting ($V = 0.05$) models.
	While the entanglement entropies take a broad range of values in both cases, with $V = 0.05$ and especially towards the middle of the spectrum, the distribution becomes strongly skewed towards higher entropies, and most of the states become volume law.
	
	However, some states retain much lower entropies, and this is especially true for the eigenstates $\ket{S_j}$ which we identified in Sec.~\ref{sec:tower} as exhibiting a large overlap with the tower of states $\ket{T_j}$.
	In particular, the $\ket{S_j}$ have very low entanglement entropies in the non-interacting case, and most retain these when interactions are turned on.
	Surprisingly, some of the $\ket{S_j}$ have even lower entropies than their approximations $\ket{T_j}$.
	On the other hand, some of the states $\ket{S_j}$ with larger $j$ do attain significantly higher entanglement entropies in the interacting case, though still small compared to most states nearby in energy: we attribute this to the large number of excitations in these states, which increases their complexity and provides more ways for the interaction $V$ to destabilize them.
	In addition to the half-chain entanglement entropy, Fig.~\ref{fig:entropy}(c)~and~(d) show $S_A$ versus subsystem size $L_A$ for the states $\ket{S_j}$.
	The data show that the entanglement of the $\ket{S_j}$ scales sub-extensively both for $V = 0$ and $V = 0.05$, except for $\ket{S_7}$ where this is not clear.
	This is in contrast to eigenstates which are nearby in energy, which obey a clear volume law.
	
	In fact, we show in Appendix~\ref{App:MPS} that the exact states $\ket{T_j}$ may be represented by a matrix-product state (MPS) of bond dimension $\chi = 4\left[\min(j+1, L/2 - j)\right]\leq L + 4$.
	Since the entanglement entropy of an MPS is at most $\log\chi$, this places a logarithmic bound on their entropy.
	We note, however, that it does not necessarily follow that the eigenstates $\ket{S_j}$ will obey this bound as well.
	\subsection{Distribution of cluster excitations}\label{sec:ClusterExpV}
	The ETH predicts that the expectation values of (local) physical operators, evaluated with respect to individual eigenstates of chaotic Hamiltonians ${\cal H}$, should form a smooth function of energy and agree with the microcanonical ensemble average for that operator.
	As a result, a distinguishing feature of ETH-violating eigenstates is a significant departure of these expectation values from the energy-resolved average.
	In the case of the ZXZ model, it is instructive to consider the expectation values of the cluster operators $K_\ell$ for the eigenstates of the model, taking particular note of the values for the eigenstates $\ket{S_j}$.
	
\begin{figure}[tb]
		\centering
		
		\includegraphics[width=0.9\columnwidth]{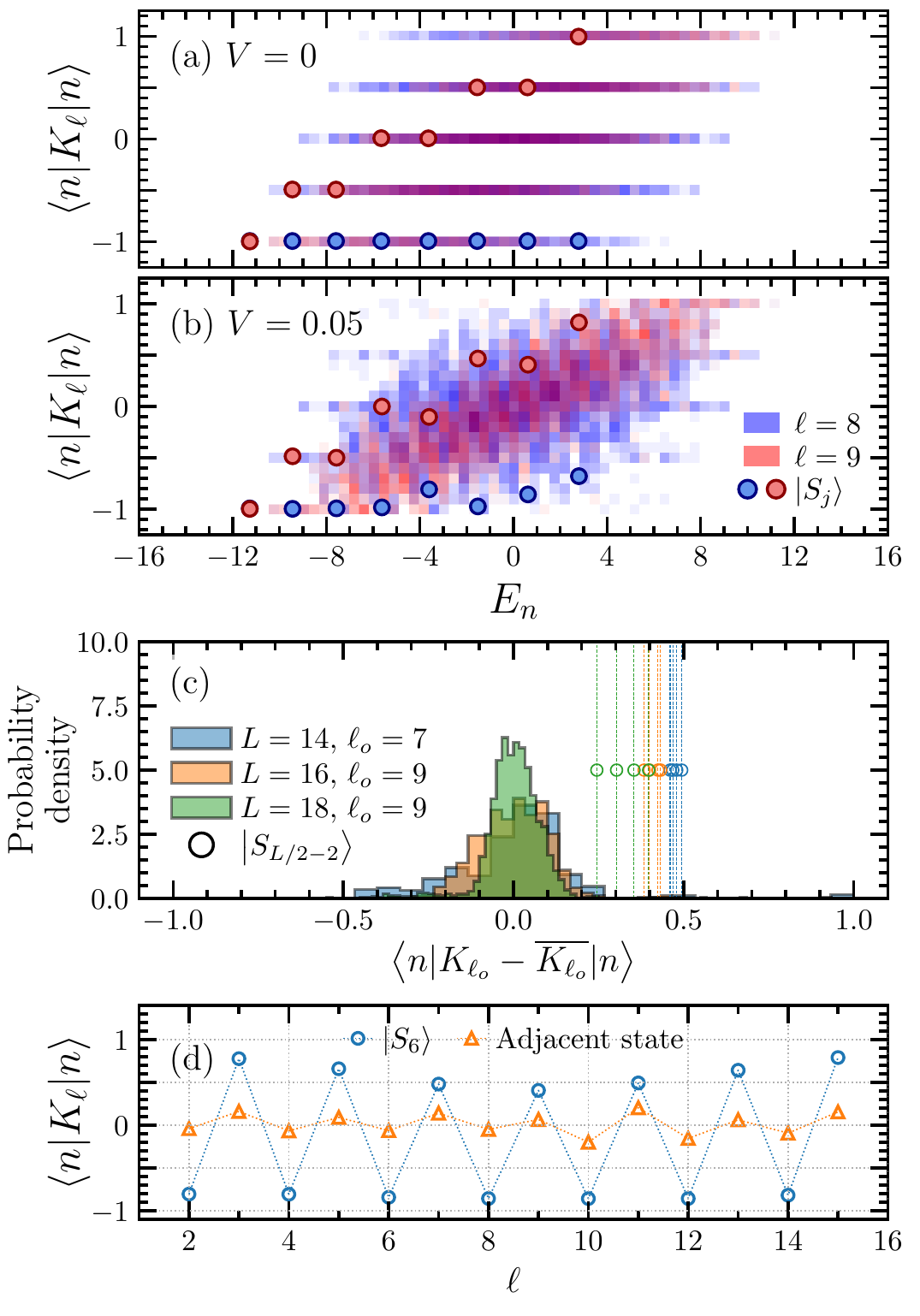}
		\caption{
			\textbf{(a)} $\expval{K_\ell}$ at sites $\ell = 8, 9$ for energy eigenstates of the non-interacting ($V = 0$) model, $L=16$, in the subspace with positive spin-flip symmetry on both sublattices. Darker colors indicate a greater number of states at that value.
			Circular markers indicate the states $\ket{S_j}$.
			\textbf{(b)} The same, but for the interacting model ($V = 0.05$).
			\textbf{(c)} Probability distribution of $\expval{K_{\ell_o}}$ for eigenstates in a narrow window around the state $\ket{S_{L/2 - 2}}$, with the data shifted such that the distributions have zero mean.
			$\ell_o = 7, 9, 9$ for $L = 14, 16, 18$ respectively, and $V = 0.05$.
			The values for the states $\ket{S_{L/2 - 2}}$ are indicated by circles, with vertical dotted lines as guides to the eye, and we include data from all four symmetry sectors.
			\textbf{(d)} $\expval{K_\ell}$ vs $\ell$, for the state $\ket{S_{L/2 - 2}}$ and a volume-law state adjacent to it in energy, for $L=16$ and $V = 0.05$.
			[$\lambda = 0.6$, $\Gamma = 0.1$ in all cases.]
		}
		\label{fig:cluster_expectation}
	\end{figure}

	Figures~\ref{fig:cluster_expectation}(a)~and~(b) show the cloud of matrix elements $\expval{K_\ell} = \matrixel{n}{K_\ell}{n}$ for two central sites in the chain with $L=16$, for the non-interacting and interacting	Hamiltonians respectively.
	The states $\ket{S_j}$ are highlighted.
	They show that in the non-interacting case, the expectation values for the central two sites fall close to a discrete set of values with little systematic dependence on energy.
	This discretization might be explained by proximity to the $\Gamma = V = 0$ point at which the $K_\ell$ are constants of motion.
	Once interactions are turned on, the distribution of $\expval{K_\ell}$ smoothens, though there is still substantial variation.
	However, some states stay close to their original values, and this is particularly prominent for the states $\ket{S_j}$ which clearly deviate from the microcanonical average.
	
	For a more refined analysis, Fig.~\ref{fig:cluster_expectation}(c) looks at the distribution of expectation values on a central odd site $\ell_o$, focusing on a narrow energy window centered around the penultimate tower state $\ket{S_{L/2-2}}$.
	This state was chosen as it is closest to zero energy for the chosen parameters.
	While the distribution appears to narrow down with increasing system size $L$, we find that the expectation values $\matrixel{S_j}{K_{\ell_o}}{S_j}$ remain distinct outliers for all $L$.
	
	The behavior of the cluster excitations is explored further in Fig.~\ref{fig:cluster_expectation}(d) by looking at their expectation values for particular states and their variation in space.
	On the one hand, the expectation value $\bra{n}	K_\ell\ket{n}$ for an eigenstate $\ket{n}$ directly adjacent to $\ket{S_j}$ is fairly uniform and thermal (as it is the case for the majority of states once interactions are turned on).
	However since eigenstates $\ket{S_j}$ are well approximated by the tower states $\ket{T_j}$, they have similar cluster expectation values.
	In particular, this means that the values between sublattices 
	differ substantially with the even sublattice being almost fully polarized.
	
	\subsection{Signatures of SPT order}
	SPT order is usually associated with properties of the ground state of a system, and in normal circumstances lost at any finite energy density.
	Despite this, it has been shown that eigenstates in a many-body localized system can remain in a sharply defined topological phase, even at infinite temperatures~\cite{Bahri2013, Huse2013, Chandran_2014}.
	More recently, it has been shown that quantum many-body scars embedded into the spectrum of a topologically ordered model can themselves have topological order, despite having a finite energy density relative to the ground state~\cite{Ok2019, Wildeboer2020, Srivatsa2020}.
	It is therefore natural to wonder whether the nonthermal states $\ket{S_j}$ discussed in this work might also retain signatures of topological order.
	To this end, we here study the so-called entanglement spectrum, defined as the eigenvalues $\varphi_\alpha$ of an ``entanglement Hamiltonian'',
	\begin{equation} \label{eq:entanglement_energy}
		H_\text{ent} = -\ln(\rho_A)\ ,
	\end{equation}
	and we choose subsystem $A$ to be the first $L/2$ sites of the system.
	
	A consequence of SPT order in the ZXZ model \eqref{eq:Ham} is that the entanglement spectrum is four-fold degenerate in the ground state, as long as $\Gamma$ and $V$ are chosen such that $\mathcal{H}$ remains in the $\mathbb{Z}_2 \times \mathbb{Z}_2$ SPT phase.
	This statement is also true of \textit{every} eigenstate in the $\Gamma = V = 0$ model.
	However for nonzero $\Gamma$, $V$, the entanglement spectrum of a state in the middle of the energy spectrum should not have this degeneracy.
	
\begin{figure}
		\centering
		\includegraphics[width=\columnwidth]{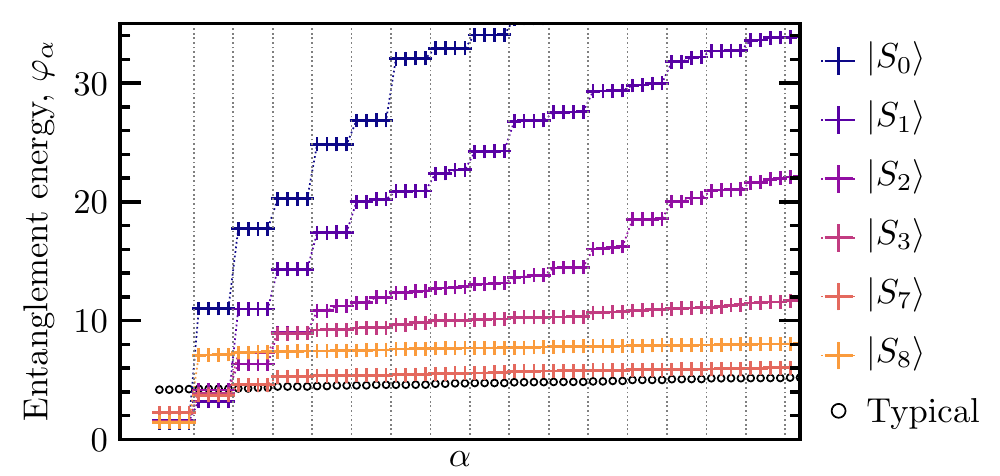}
		\caption{
			Ordered ``entanglement energies'' $\varphi_\alpha$ [cf.\ Eq.~\eqref{eq:entanglement_energy}] of selected exemplary states in $\left\{\ket{S_j}\right\}$.
			For comparison, we also show the entanglement spectrum of a typical eigenstate close to zero energy.
			We have for $L = 18$, $\Gamma = 0.1$, $V = 0.05$ in all cases.
		}
		\label{fig:entanglement_spectrum}
	\end{figure}

	In spite of this, we find signs of this degeneracy for all of the states in the tower, $\ket{S_j}$.
	Figure~\ref{fig:entanglement_spectrum} gives illustrative examples, as follows.
	As $\ket{S_0}$ is of course also the ground state, the four-fold degeneracy is clear throughout the entanglement spectrum, and there is a large gap between the ``ground'' quadruplet and the next set of values.
	This serves as a benchmark for the behavior of the other $\ket{S_j}$.
	States $\ket{S_1}$ and $\ket{S_2}$ still retain a four-fold degeneracy, though with smaller gaps, and a slight breakdown of this degeneracy at higher entanglement energies.
	Moreover, even though $\ket{S_7}$ is close to the middle of the spectrum, the first dozen $\varphi_\alpha$ clearly form well-separated quadruplets.
	Finally even the highest state in the tower, $\ket{S_8}$ keeps the degeneracy in the first quadruplet with a gap almost comparable to that of the ground state $\ket{S_0}$.
	This may be a signature that these nonthermal states retain SPT order.
	In particular, the entanglement spectrum of the $\ket{S_j}$ is 
	in stark contrast to a typical state in the spectrum (in this case, chosen to 
	have close to zero energy), which shows no signs of the four-fold degeneracy.
	
	We leave it to future work to study this finding in more detail, e.g., by looking at other indicators such as the topological entanglement entropy or the appropriate string order parameter~\cite{Lavasani_2021}.
	
	\section{Nonequilibrium dynamics of the ZXZ model}\label{sec:Dynamics}
	
	Having established the presence of scarred eigenstates throughout the 
	spectrum of ${\cal H}$, we now turn to the dynamical properties of the ZXZ 
	model.
	Given the nonthermal character of the states $\ket{S_j}$, combined with their signatures of topological order, it is plausible that the dynamics of ${\cal H}$ shows anomalous behavior as well.
	In contrast to Ref.~\cite{Kemp2019}, where the focus was on the long-lived edge mode, here we mainly scrutinize the dynamics in the bulk of the system.
	
	\subsection{Infinite-temperature dynamics}
	
\begin{figure}[tb]
		\centering
		\includegraphics[width = 1\columnwidth]{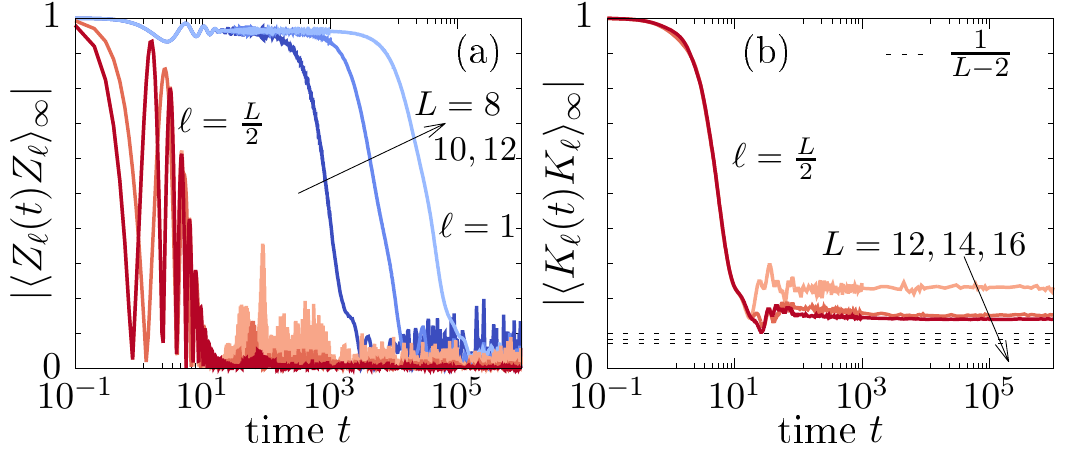}
		\caption{
			Infinite-temperature autocorrelation functions
			[Eq.~\eqref{eq:Infty}] at the
			edge
			and in the bulk for (a) $Z_\ell$ and (b)
			$K_\ell$.
			Data is obtained by exact diagonalization for different system sizes $L$ as indicated by the arrows.
			Note that data in (a) is
			analogous to Ref.~\cite{Kemp2019}.
			The dashed lines in (b) signal the equipartition value $1/(L-2)$.
			The other parameters are chosen as $V = 0.05$, $\Gamma = 0.1$, and $\lambda = 0.6$.
		}
		\label{FigEdgeMode}
	\end{figure}

	As a starting point, Figs.~\ref{FigEdgeMode}(a)~and~(b) show the infinite-temperature autocorrelation functions $|\langle A_\ell(t) A_\ell \rangle_\infty|$ of local spin ($A_\ell = Z_\ell$) or cluster operators ($A_\ell = K_\ell$) respectively,
	\begin{equation}
		\label{eq:Infty} \langle A_\ell(t) A_\ell \rangle_\infty = \frac{\text{Tr}[A_\ell(t) A_\ell]}{2^L}\ ,
	\end{equation}
	where $A(t) = e^{i{\cal H}t}Ae^{-i{\cal H}t}$.
	Specifically, Fig.~\ref{FigEdgeMode}(a) shows $|\langle Z_\ell(t) Z_\ell \rangle_\infty|$ for sites $\ell = 1$ and $\ell = L/2$, i.e., at the edge and in the bulk of ${\cal H}$.
	Setting $\lambda = 0.6$ and plotting results for different system sizes $L$, we find that $|\langle Z_\ell(t) Z_\ell \rangle_\infty|$ quickly decays towards zero for $\ell = L/2$ on a time scale which is essentially independent of $L$.
	In contrast, for $\ell = 1$, $|\langle Z_\ell(t) Z_\ell \rangle_\infty|$ only starts to decay after a time that increases exponentially with $L$, illustrating the intriguing result of Ref.~\cite{Kemp2019} that ${\cal H}$ hosts a long-lived edge mode which is stable even at infinite temperature if the model \eqref{eq:Ham} is dimerized with $\lambda \neq 1$.
	
	Contrary to $|\langle Z_\ell(t) Z_\ell \rangle_\infty|$, the autocorrelation function $|\langle K_\ell(t) K_\ell \rangle_\infty|$ of local cluster operators does not exhibit such a distinct dependence on the choice of the site $\ell$.
	In Fig.~\ref{FigEdgeMode}(b), we exemplarily show $|\langle K_\ell(t) K_\ell \rangle_\infty|$ for $\ell = L/2$ which is found to decay rapidly to a constant long-time value (the dynamics for other $\ell$ are very similar).
	Interestingly, this asymptotic long-time value is nonzero [in contrast to the data for $Z_\ell$ in panel (a)] and also clearly larger than $1/(L-2)$ [dashed lines in Fig.~\ref{FigEdgeMode}(b)], which would have indicated that the single cluster excitation has uniformly spread over the whole system.
	While the long-time value is a direct consequence of the broad distribution of $\langle n|K_\ell|n \rangle$ shown in Fig.~\ref{fig:cluster_expectation}, we note that $|\langle K_\ell(t\to\infty) K_\ell \rangle_\infty|$ seems to decrease slightly with increasing $L$.
	This could indicate that the fraction of scarred low-entangled 
	eigenstates in the spectrum of $\cal{H}$ (which were found to yield rather 
	extremal values of $\langle n|K_\ell|n \rangle$) becomes smaller when 
	approaching the thermodynamic limit $L \to \infty$.
	
	In the following, we argue that the apparent nonthermal long-time value of $\langle K_\ell(t) K_\ell \rangle_\infty$ can be understood as a consequence of the stability of local cluster expectation values $\langle \psi|K_\ell(t)|\psi\rangle$ for specific out-of-equilibrium initial states $\ket{\psi}$ chosen from the cluster basis.
	In particular, we show that this stability is related to
	the dimerization of the ZXZ model with $\lambda = 0.6$,
	which causes cluster excitations to remain essentially confined to one of
	the two sublattices.
	\subsection{Dynamics of cluster-basis initial states}
	
	To proceed, we study the dynamics of initial states $\ket{\psi(0)}$ drawn from the cluster basis, i.e., eigenstates of the clean ZXZ model with $\Gamma = V = 0$.
	At time $t = 0$, such states therefore yield definite expectation values $\pm 1$ for local cluster operators,
	\begin{equation}
		\bra{\psi(0)} K_\ell \ket{\psi(0)} = \pm 1\ ,\quad 2\leq \ell \leq L-1\ .
	\end{equation}
	Under evolution with respect to ${\cal H}$ with $\Gamma, V > 0$, however, the expectation values $\bra{\psi(t)} K_\ell \ket{\psi(t)}$ are not conserved and can decay with time.
	We here investigate the dependence of the resulting dynamics on the choice of the particular initial state $\ket{\psi(0)}$ and the dimerization parameter $\lambda$.
	
	This analysis provides a phenomenological picture of the 
	obstruction to thermalization in this model.  We investigate the scenario where 
	two subsystems (odd and even sublattices) are prepared at different energy 
	densities and study the relaxation between them. The dynamics of the cluster 
	excitations between the odd and even sublattices are suppressed where the 
	relaxation time between the two subsystems even appears to diverge for 
	dimerization parameters deviating from $1$. We show that when the even 
	sublattice is kept at zero temperature, although the odd sublattice is highly 
	excited, the dynamics only leads to thermalization within the sublattices but 
	not between them thus providing a novel feature where a finite temperature 
	thermalizing system is unable to heat up a zero temperature state.
	
\begin{figure}[tb]
		\centering
		\includegraphics[width =1\columnwidth]{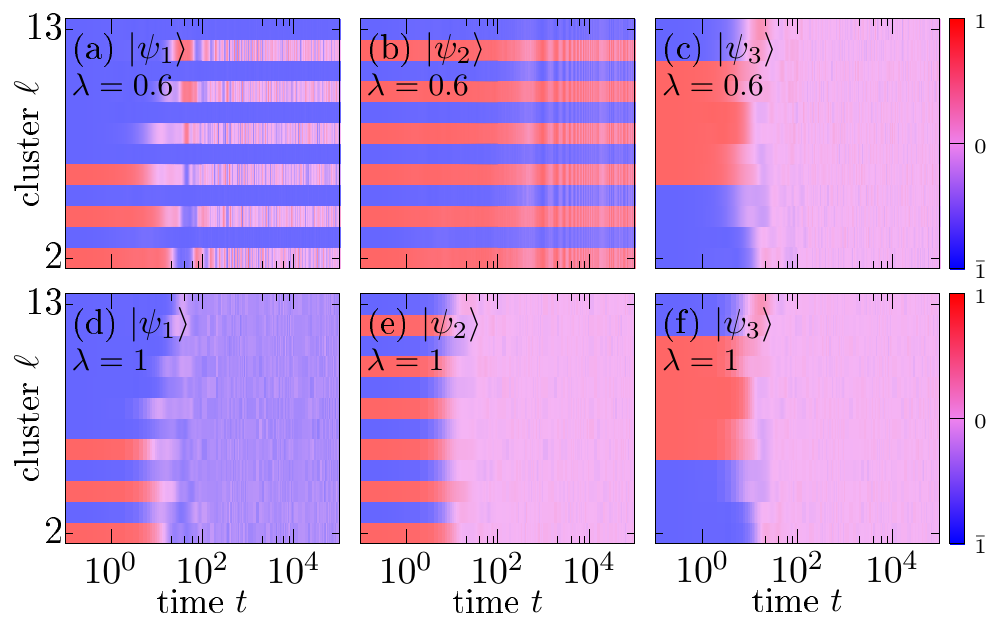}
		\caption{
			Quench dynamics $\langle 
			K_\ell(t)\rangle_{\ket{\psi}} = \bra{\psi(t)} K_\ell \ket{\psi(t)}$ for three 
			exemplary states $\ket{\psi_1}$-$\ket{\psi_3}$ from the cluster basis.
			Panels (a)-(c) show data for the dimerized model 
			with $\lambda = 0.6$, while panels (d)-(f) show data for $\lambda = 1$.
			\textbf{[(a),(d)]}
			Three excitations on even sublattice, 
			$\ket{\psi_1} = |1\bar{1}1\bar{1}1\bar{1}\bar{1}\cdots\rangle$; 
			\textbf{[(b),(e)]} Fully excited even sublattice, $\ket{\psi_2} = 
			|1\bar{1}1\bar{1}\cdots\rangle$; \textbf{[(c),(f)]} Excitations on both 
			sublattices, $\ket{\psi_3} = 
			|\bar{1}\bar{1}\bar{1}\bar{1}111111\bar{1}\bar{1}\rangle$.
			We have $\Gamma = 0.1, V = 0.05$ and $L = 14$ in 
			all cases.
		}
		\label{FigProfile}
	\end{figure}

	In Fig.~\ref{FigProfile}, we study the quench dynamics $\langle K_\ell(t)\rangle_{\ket{\psi}}$ for three exemplary initial states $\ket{\psi_1}$-$\ket{\psi_3}$,
	\begin{equation}
		\label{eq:QuenchDy}
		\expval{K_\ell(t)}_{\ket{\psi}} = \mel{\psi(t)}{K_\ell}{\psi(t)}\ ,
	\end{equation}
	where we particularly compare the dynamics for $\lambda = 0.6$ [Figs.~\ref{FigProfile}(a)-(c)] and $\lambda = 1$ [Figs.~\ref{FigProfile}(d)-(f)].
	The choice of the states
	$\ket{\psi_1}-\ket{\psi_3}$ is motivated by the
	construction of the tower
	of states $\ket{T_j}$ (and the corresponding nonthermal
	eigenstates
	$|S_j\rangle$) in Sec.~\ref{sec:tower} above.
	Specifically, in the case of $\ket{\psi_1}$
	and $\ket{\psi_2}$, the clusters of the odd
	sublattice are all in their ground state, while cluster excitations are present on
	the even sublattice.
	In contrast, the state $\ket{\psi_3}$ features a finite number of cluster excitations on both sublattices.
	
	The data in Fig.~\ref{FigProfile} exemplifies
	a strong
	dependence of $\langle K_\ell(t) \rangle_{\ket{\psi}} =
	\bra{\psi(t)}
	K_\ell \ket{\psi(t)}$ on the choice of $\lambda$.
	On the one hand, in the case of $\lambda = 0.6$,  we observe that even though the cluster excitations spread through the system, they do so by remaining almost perfectly confined to the even sublattice.
	This is especially clear for the state $\ket{\psi_1}$ in Fig.~\ref{FigProfile}(a).
	As a consequence, the initial inhomogeneous cluster configuration does not equilibrate, such that thermalization of the full system is prevented even on the very long time scales $t \leq 10^5$ shown here.
	This is in contrast to the state $\ket{\psi_3}$ in Fig.~\ref{FigProfile}(c), which appears to thermalize very rapidly.
	As cluster excitations are already present on both sublattices, the effective restriction of the cluster excitations to one sublattice does not have a strong effect in this case.
	Let us note that such a discrepancy between initial states with or without excitations on both sublattices has been already exemplified in terms of the Loschmidt echo ${\cal L}(t)$ in Fig.~\ref{fig:chain_diagram}(d).

	On the other hand, in the case of $\lambda = 1$, we find that especially for the states
	$\ket{\psi_1}$ and $\ket{\psi_2}$,
	$\langle K_\ell(t) \rangle_{\ket{\psi}}$ behaves very
	differently.
	In particular, the initial cluster profile is found to spread very rapidly through the whole system, i.e., the cluster excitations can move freely onto the other sublattice as well.
	This strong dependence of the quench dynamics on the 
	dimerization parameter $\lambda$ is an important result of the present paper.
	Moreover, connecting the dynamics in Fig.~\ref{FigProfile} to
	the eigenstate properties
	discussed above in Sec.~\ref{sec:entropy}, let us
	mention that the states
	$\ket{\psi_1}$ and $\ket{\psi_2}$ appear to overlap
	dominantly with a sub-volume law
	entangled eigenstate, while most of the spectral weight
	of
	$\ket{\psi_3}$ is on a volume-law entangled eigenstate
	instead (see Appendix
	\ref{App::Overlap} for
	details).
	
\begin{figure}[tb]
		\centering
		\includegraphics[width = 1\columnwidth]{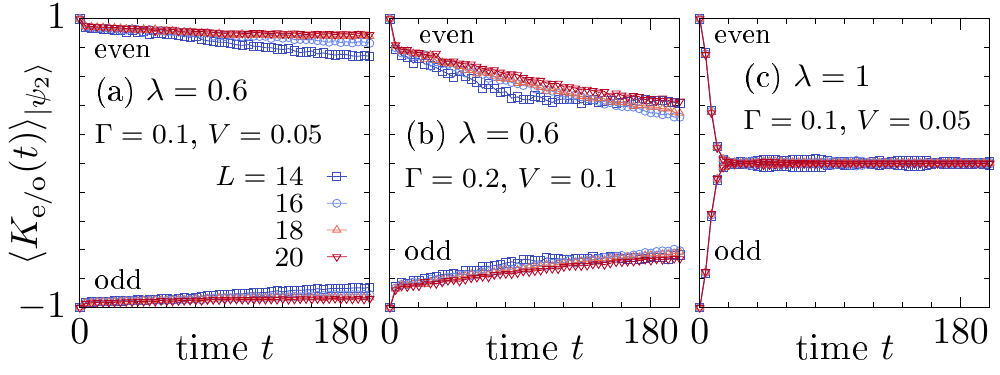}
		\caption{
			Averaged cluster expectation
			value $\langle K_\text{e/o}(t)\rangle_{\ket{\psi_2}}$ on even/odd sublattice [Eq.~\eqref{eq:EvenOdd}] for the initial state $\ket{\psi_2}$ [see Figs.~\ref{FigProfile}(b),(e)].
			Data is shown for system sizes $L = 14,16,18,20$.
			In panels (a) and (b), we show data for the dimerized model with $\lambda = 0.6$, while panel (c) shows data for $\lambda = 1$.
			In (a) and (c), we have $\Gamma = 0.1$ and $V = 0.05$.
			Panel (b) considers slightly stronger perturbations $\Gamma = 0.2$ and $V = 0.1$.
		}
		\label{fig:Fig8}
	\end{figure}

	In order to analyze the quench dynamics in
	more detail,
	we now focus on the initial state $\ket{\psi_2}$ from
	Figs.~\ref{FigProfile}(b) and (e).
	Specifically, in Fig.~\ref{fig:Fig8}, we show the averaged cluster expectation value $\langle K_\text{e/o}(t)\rangle_{\ket{\psi}}$ on the even~(e) and odd~(o) sublattice,
	\begin{equation}
		\label{eq:EvenOdd} \langle K_\text{e}(t)\rangle_{\ket{\psi}} = \frac{2}{L-2}\sum_{\ell = 1}^{L/2-1} \langle \psi(t)|K_{2\ell}|\psi(t)\rangle\ ,
	\end{equation}
	where $K_{2\ell} \to K_{2\ell +1}$ in the case of $\langle K_\text{o}(t)\rangle_{\ket{\psi}}$.
	Note that in Fig.~\ref{fig:Fig8}(b),
	we now also include data for slightly larger values of
	the perturbations
	$\Gamma = 0.2$ and $V = 0.1$.
	
	As already expected given Fig.~\ref{FigProfile}, the data in Figs.~\ref{fig:Fig8}(a)-(c) confirm that $\langle K_\text{e/o}(t)\rangle_{\ket{\psi}}$ clearly depends on the choice of
	$\lambda$.
	In the case of $\lambda = 0.6$ [Figs.~\ref{fig:Fig8}(a),(b)], the
	dynamics of $\langle K_\text{e/o}(t)\rangle_{\ket{\psi}}$ is very
	slow (although it is somewhat faster for the larger values of
	$\Gamma,V$) and the even
	and odd sublattices do not equilibrate on the time
	scales $t \leq 200$ shown here.
	Interestingly, comparing data for different system sizes $L = 14-20$, it appears that the dynamics becomes even slower for increasing $L$.
	While it is difficult to decide with numerical means whether $\langle K_\text{e/o}(t)\rangle_{\ket{\psi}}$ will eventually thermalize for $L \to \infty$ and $t \to \infty$ [although we note that the cluster profiles in Figs.~\ref{FigProfile}(a) and (b) stay	nonthermal even for $t > 10^4$], the results in Fig.~\ref{fig:Fig8}
	strongly suggest that the time scale of (potential) thermalization is
	significantly longer in the dimerized model with
	$\lambda = 0.6$ (even for the larger values of $\Gamma$ and $V$).
	In particular, as shown in Fig.~\ref{fig:Fig8}(c), $\langle
	K_\text{e/o}(t)\rangle_{\ket{\psi}}$ quickly
	approaches zero for $\lambda = 1$
	on a $L$-independent time scale $t \lesssim 10$.
	Thus, by virtue of the dimerization parameter $\lambda\neq 1$ 
	(i.e.\ in the same parameter regime where Ref.~\cite{Kemp2019} reported the 
	long-lived edge mode) it is possible to induce dynamics with long coherence 
	times in the bulk degrees of freedom of the ZXZ model \eqref{eq:Ham}, at least when looking at 
	appropriate operators and initial states.

	Having exemplified the dynamics of local cluster operators $K_\ell$ in Fig.~\ref{FigProfile} for the three states $\ket{\psi_1}$ - $\ket{\psi_3}$, we now study the quench dynamics for a wider class of initial states from the cluster basis.
	In this context, a useful quantity to probe the ``stability'' of the initial cluster distribution for some state $\ket{\psi}$ is the time- and space-averaged correlation function $\overline{K_{\ket{\psi}}(t)}$ (c.f.\ Ref.~\cite{Pancotti_2020}),
	\begin{equation}
		\label{eq:KK}
		\overline{K_{\ket{\psi}}(t)} = \frac{1}{L-2}
		\sum_{\ell=2}^{L-1}
		\frac{1}{t}
		\int_0^{t}
		\langle \psi|K_\ell(\tau)K_\ell \ket{\psi} d\tau\ .
	\end{equation}
	If the long-time value of $\overline{K_{\ket{\psi}}(t)}$ remains nonzero or even relatively close to $\overline{K_{\ket{\psi}}(t\to \infty)} \approx 1$, this can be interpreted as an indication that at least some of the local cluster information is preserved.
	
\begin{figure}[tb]
		\centering
		\includegraphics[width =1\columnwidth]{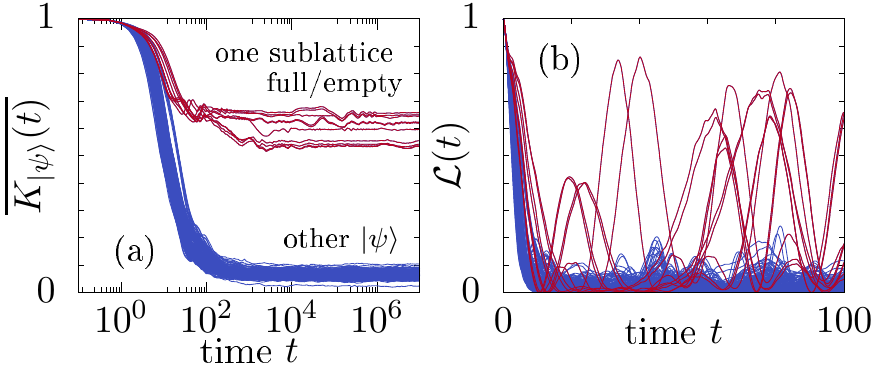}
		\caption{
			\textbf{(a)}
			Time- and space-averaged correlation function $\overline{K_{\ket{\psi}}(t)}$ and \textbf{(b)} Loschmidt echo ${\cal L}(t) = |\langle \psi(t)|\psi\rangle|^2$, for the dimerized model with $\lambda = 0.6$ and fixed system size $L = 12$.
			Data is shown those states $\ket{\psi}$ from the cluster basis with an energy $E_{\ket{\psi}} \in [-0.5,0.5]$ in the center of the spectrum.
			Initial states where one of the two sublattices is fully excited/fully in the ground state (red curves) saturate to high values of $\overline{K_{\ket{\psi}}(t)}$ and exhibit distinct oscillations in ${\cal L}(t)$.
		}
		\label{Fig8}
	\end{figure}

	While there are $2^{L-2}$ different initial states $\ket{\psi}$ in a given symmetry sector (as there are $L-2$ cluster sites), we here consider only those states which have a mean energy $E_{\ket{\psi}}$ relatively close to the center of the spectrum,
	\begin{equation}
		-0.5 \leq E_{\ket{\psi}} \leq 0.5\ ,\quad E_{\ket{\psi}} = \bra{\psi}{\cal H}\ket{\psi}\ .
	\end{equation}
	For a thermalizing system, one would expect that initial states with approximately the same energy will yield very similar outcomes of local observables at long times.
	In Fig.~\ref{Fig8}(a), we find that this expectation is clearly violated for the dimerized model with $\lambda = 0.6$.
	In particular, while the time- and space-averaged correlator $\overline{K_{\ket{\psi}}(t)}$ decays towards zero for the majority of initial states, we identify a rare number of $\ket{\psi}$ where $\overline{K_{\ket{\psi}}(t\to \infty)}$ is clearly nonzero.
	Upon inspection, it turns out that these rare $\ket{\psi}$ are exactly those states where the clusters of one sublattice are all either excited or in the ground state.
	Note that this finding is perfectly consistent with our previous results from Fig.~\ref{FigProfile}(a)-(c).
	Namely, as cluster excitations remain essentially confined to their original sublattice, inhomogeneous cluster distributions cannot equilibrate over the whole system, such that thermalization of the full system is impeded.
	Consequently, for initial states $\ket{\psi}$ where cluster excitations are present only on one sublattice, the cluster expectation values on the other sublattice remain almost unchanged and one obtains a saturation value $\overline{K_{\ket{\psi}}(t\to \infty)} \geq 0.5$.
	
	The nonthermal dynamics of the local cluster operators $K_\ell$ for some of the states $\ket{\psi}$ also reflects itself in the Loschmidt echo ${\cal L}(t) = |\langle \psi(t)|\psi\rangle|^2$, which is shown in Fig.~\ref{Fig8}(b).
	Focusing again on those states with $E_{\ket{\psi}} \in [-0.5,0.5]$, we find that ${\cal L}(t)$ quickly decays and stays close to zero for the majority of $\ket{\psi}$.
	However, once again, we observe that those states where the clusters of one sublattice are all either excited or in the ground state behave drastically different.
	Specifically, these rare $\ket{\psi}$ also decay towards zero initially, but exhibit distinct revivals of ${\cal L}(t)$ at short to intermediate times.
	These pronounced revivals can be understood as a direct consequence of the restricted mobility of cluster excitations for $\lambda = 0.6$.
	Specifically, the detuning of the two sublattices causes states $\ket{\psi}$ with cluster excitations only on one sublattice to be relatively weakly connected to other states with excitations on both sublattices.
	As a consequence, such initial states explore only a smaller part of the Hilbert space, and local memory of the initial state (at least on one of the two sublattices) is preserved even at long times.
	While not shown here, we have checked that such strong revivals 
	of ${\cal L}(t)$ are absent in the nondimerized model with $\lambda = 1$.

	\section{Discussion}\label{sec:Discussion}
	
	We have studied the eigenstate properties and the dynamics of an interacting spin-$1/2$ chain with three-site cluster terms, hosting a symmetry-protected topological phase.
	We have particularly focused on the weakly interacting regime in the presence of a dimerization parameter $\lambda \neq 1$, which energetically detunes the clusters on odd and even sites.
	This region in parameter space has been recently identified in Ref.~\cite{Kemp2019} to stabilize a long-lived edge mode even at infinite temperature.
	In the present work, we have scrutinized this intriguing parameter regime with respect to quantum many-body scars and the out-of-equilibrium dynamics of the bulk degrees of freedom in the bulk of the SPT phase.
	
	Summarizing our main results, we have shown that there is a coexistence of volume-law and area-law entangled eigenstates throughout the spectrum of ${\cal H}$.
	The latter are akin to quantum many-body scars in the sense that an equidistant ``tower'' of such states can be constructed by the repeated application of a certain type of ``raising'' operator.
	In the present case, this construction consists of coherent cluster excitations on one of the two sublattices with the other sublattice remaining in the ground state.
	Furthermore, we have provided a detailed characterization of the 
	properties of the scarred eigenstates and demonstrated that they feature 
	nonthermal expectation values of local cluster operators in the bulk and 
	remarkably exhibit signatures of topological order even at finite temperature.
	In contrast to the usual phenomenology of quantum scars, where the scar states are distinct outliers in an otherwise strongly thermalizing spectrum with extensive entanglement, the ZXZ models appears to be less chaotic, e.g., in the sense that a large number of eigenstates exhibits comparatively low entanglement.
	This fact might be partially attributed to the definition of ${\cal H}$ with open boundary conditions and the presence of edge modes, as well as to the integrability-breaking interactions being comparatively weak.
	In fact, as shown in Appendix \ref{App::Chaos}, the level spacing 
	distribution of ${\cal H}$ indicates that the ZXZ model for weak interactions 
	actually resides in an intermediate regime between integrability and full 
	quantum chaos, although we here cannot rule out the impact of finite-size 
	effects.
	
	In addition to the properties of eigenstates, we have studied the out-of-equilibrium dynamics of ${\cal H}$.
	We particularly considered quantum quenches starting from eigenstates of the noninteracting model which have definite eigenvalues of local cluster operators, $\langle \psi(0)|K_\ell\ket{\psi(0)} = \pm 1$.
	We have shown that the dynamics of the clusters drastically depends on the choice of the dimerization parameter $\lambda$.
	Specifically, we have found that a finite detuning (here $\lambda = 0.6$) effectively restricts the mobility of cluster excitations to within each of the two sublattices, such that inhomogeneous cluster distributions cannot equilibrate over the whole system.
	As a consequence, especially for those initial states where cluster excitations are present on only one sublattice, the cluster expectation values of the other sublattice remain almost unchanged even on very long time scales.
	This nonthermal dynamics is in stark contrast to the nondimerized model ($\lambda = 1$), where the two sublattices are found to mix rapidly and thermalization is observed.
	Thus, in the same parameter regime where Ref.~\cite{Kemp2019} has found robust edge modes at infinite-temperature, the bulk dynamics likewise exhibits long coherence times when considering appropriate operators and initial states.
	Let us note, however, that the occurrence of nonthermal cluster dynamics in the bulk does not necessarily require the presence of a long-lived edge mode.
	In particular, we have checked that a phenomenology similar to Fig.~\ref{FigProfile} holds for periodic boundary conditions as well.
	
	Our work raises a number of questions.
	While the discussion of quantum many-body scars is usually concerned with strongly thermalizing models where all but a few eigenstates obey the ETH, it might be interesting to extend the notion of scars to quantum integrable models as well.
	In this work, we have shown that the noninteracting version of the ZXZ model (corresponding to two uncoupled Ising chains) also hosts a number of area-law entangled eigenstates, some of which appear to remain stable for finite interactions.
	On the one hand, it would be interesting to better understand the connection between such states and the well-known local conservation laws in the transverse-field Ising model with open boundaries \cite{Fagotti_2016}, as well as the stability of such conservation laws against perturbations.
	In this context, let us note that connections between quantum many-body scars and the proximity to integrable points have already been discussed in Refs.~\cite{Khemani_2019, Desaules2021}.
	On the other hand, thermalization in integrable models is usually understood with respect to a suitable generalized Gibbs ensemble (GGE), which accounts for the extensive number of conservation laws \cite{Vidmar_2016, Essler_2016}.
	Analogous to chaotic models, one might speculate that the presence of ``scars'' in integrable models can prevent thermalization to a GGE for certain out-of-equilibrium states which exhibit a dominant overlap with such eigenstates.
	
	The role of disorder on the topological nature of quantum scars in the 
	ZXZ model is a question of interest in the context of many-body localization.
	It has been shown for example that the PXP model can be localized \cite{Chen2018}, but the constraints in the model also induce interactions which in certain cases prevents MBL \cite{Sierant2021}. 
	Moreover, relatively weak disorder may destroy scars \cite{MondragonShem2020}, even though at strong disorder 
	many-body localization reinstates the topological edge modes. It would be 
	interesting to study the phase transition between the various excited state 
	phases with topological features in intermediate regimes between scars and 
	localization.
	
	\subsection*{Acknowledgements}
	This work was funded by the European Research Council (ERC) under the European Union's Horizon 2020 research and innovation programme (Grant agreement No.~853368). 
	J.J.\ is supported by a UK Engineering and Physical Sciences Research Council (EPSRC) studentship (Project Ref.~2252612) under the EPSRC Doctoral Training Partnership (DTP) with University College London (Grant Ref.~EP/R513143/1).
	\FloatBarrier
	\appendix
	\section{Eigenstates of the clean ZXZ model (\texorpdfstring{$\mathbf{\Gamma = V = 0}$}{Gamma = V = 0})}\label{app::cluster_basis}
	There exists a set of states $\ket*{\psi_{\{\pm_\ell\}}^{g_\mathrm{e}, g_\mathrm{o}}}$ which are mutual eigenstates of the cluster operators and have definite $\mathbb{Z}_2 \cross \mathbb{Z}_2$ symmetry values, forming a complete basis within each symmetry sector.
	
	Let us define the state $\ket*{\psi_{\{\pm_\ell\}}^{g_\mathrm{e}, g_\mathrm{o}}}$ such that,
	\begin{align}
		\hat{G}_\mathrm{e/o} \ket{\psi_{\{\pm_\ell\}}^{g_\mathrm{e}, g_\mathrm{o}}} & = g_\mathrm{e/o} \ket{\psi_{\{\pm_\ell\}}^{g_\mathrm{e}, g_\mathrm{o}}}, \label{eqn:cluster1} \\ K_\ell \ket{\psi_{\{\pm_\ell\}}^{g_\mathrm{e}, g_\mathrm{o}}} & = \pm_\ell \ket{\psi_{\{\pm_\ell\}}^{g_\mathrm{e}, g_\mathrm{o}}}\ , \quad 2 \leq \ell \leq L - 1\label{eqn:cluster_action_z}\ ,
	\end{align}%
	where $\hat{G}_\mathrm{e/o}$ are the symmetry operators for even and odd site spin-flip symmetry respectively, and $g_\mathrm{e/o} = \pm1$ their eigenvalues.
	Moreover, $\lbrace \pm_\ell \rbrace$ is here used as
	an abbreviation for the local cluster expectation values being
	either $+1$ or $-1$ on each site $\ell$.
	Equations \eqref{eqn:cluster1} and
	\eqref{eqn:cluster_action_z} do not uniquely specify the phase
	of these states, but they can be defined concretely for our purposes as,
	\begin{gather}
		\label{eq:cluster_definition}
		\ket{\psi_{\{\pm_\ell\}}^{g_\mathrm{e}, g_\mathrm{o}}} =
		\sqrt{2^L} \left(\prod_{\ell=2}^{L-1}
		K_\ell^{\pm_\ell}\right) P_\mathrm{e}^{g'_\mathrm{e}} P_\mathrm{o}^{g'_\mathrm{o}} \ket{\uparrow}^{\otimes L}\ ,
	\end{gather}%
	where $P_\mathrm{e/o}^\pm = (1 \pm \hat{G}_\mathrm{e/o})/2$, and $g'_{\mathrm{e/o}} = g_{\mathrm{e/o}}^{L/2 - 1} = \pm1$.
	The state $\ket{\psi_{\bar{1}/0}}$ considered in the
	context of Fig.~\ref{fig:tower} is constructed by applying
	lowering operators only on the even sites, and adjusting the normalization accordingly.
	
	It is easy to check that the states $\ket*{\psi_{\{\pm_\ell\}}^{g_\mathrm{e}, g_\mathrm{o}}}$ have the correct symmetry eigenvalues, after noting that $\acomm*{K_{2\ell}^\pm}{G_e} = \comm*{K_{2\ell}^\pm}{G_o} = 0$ and likewise $\comm*{K_{2\ell + 1}^\pm}{G_e} = \acomm*{K_{2\ell + 1}^\pm}{G_o} = 0$.
	Noting also that $K_\ell K^\pm_\ell = \pm K^\pm_\ell$ shows that they yield the correct cluster-operator eigenvalues.
	
	It is simple to show that these states are mutually orthogonal.
	Take two states which differ  in their cluster eigenvalues at at least one site.
	Since $(K^\pm)^\dagger = K^\mp$, taking the inner product of these states, substituting in \eqref{eq:cluster_definition} and grouping the operators by site (using the fact that operators on different sites commute) will lead to at least one factor $(K^\pm)^2 = 0$.
	Hence the inner product must vanish.
	It is clear also that states with different symmetry eigenvalues must be orthogonal.
	
	Since there are $L-2$ sites each with two choices for cluster eigenvalues, and also four symmetry sectors, there are $4 \times 2^{L-2} = 2^L$ cluster states -- exactly the number of states in a system of size $L$.
	Hence this is a complete, orthonormal basis.
	\section{Onset of chaos}\label{App::Chaos}
	A useful indicator of whether a system is integrable or chaotic is given by the ratio of adjacent level spacings $\tilde{r}$~\cite{Huse2007, Atas2013},
	\begin{equation}
		\label{eq:rtilde} \tilde{r} = 
		\frac{\text{min}\lbrace\Delta_n,\Delta_{n+1}\rbrace}{\text{max}\lbrace\Delta_n, \Delta_{n+1}\rbrace} ,
	\end{equation}
	where $\Delta_n = E_{n+1} - E_n$ is the spacing between consecutive energy levels.
	On the one hand, for an integrable system, the energy levels follow a Poisson distribution~\cite{Berry1977}.
	On the other hand, for a nonintegrable ``chaotic'' system, the levels will follow a similar distribution to that of a random matrix drawn from a Gaussian ensemble~\cite{Bohigas1984, Haake2001}, e.g., the Gaussian orthogonal ensemble (GOE) in case of ${\cal H}$ having real entries.
	Such ``chaotic'' systems are expected to obey the ETH \cite{D_Alessio_2016, Borgonovi_2016}.
	
\begin{figure}[tb]
		\centering
		\includegraphics[width=0.95\columnwidth]{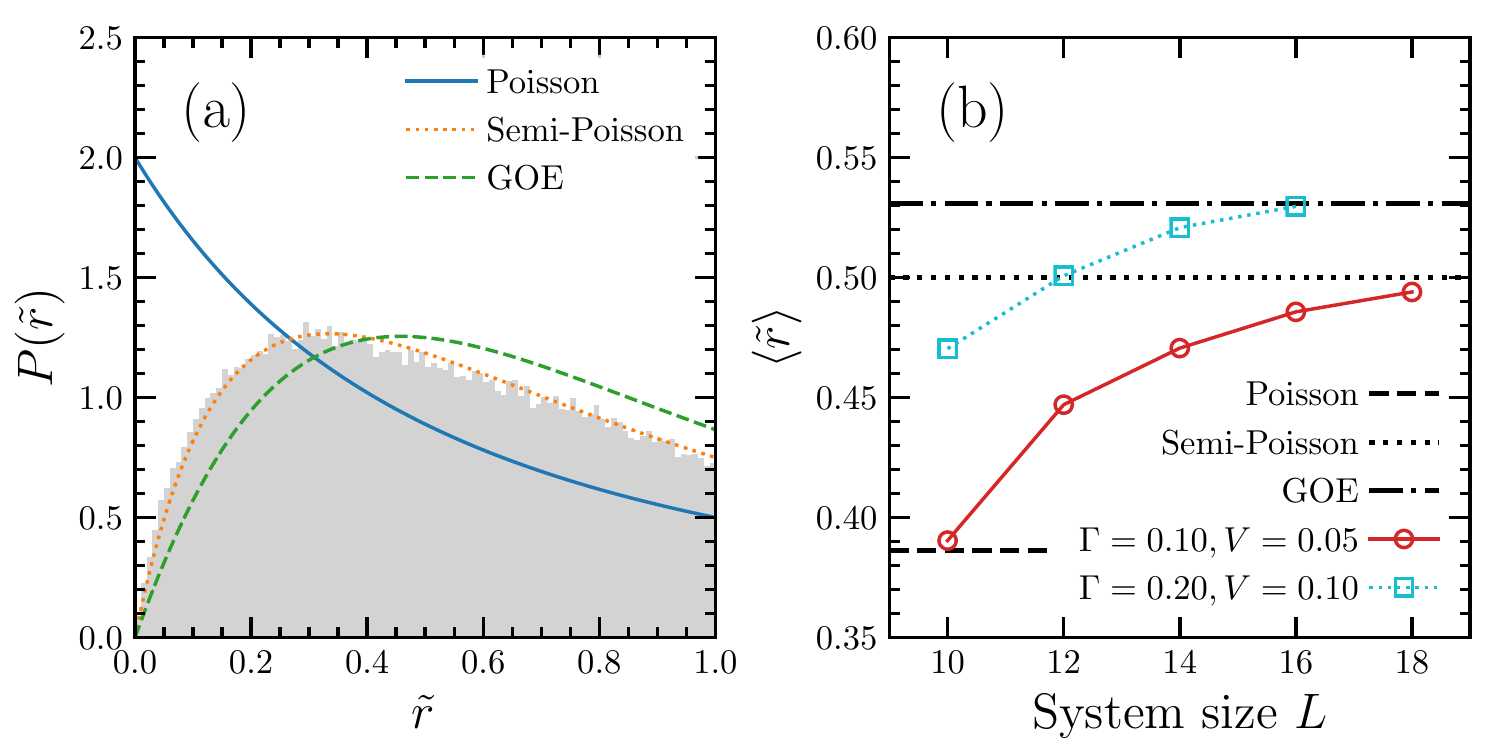}
		\caption{
			\textbf{(a)}
			Distribution of $\tilde{r}$ for $L=18$, $\Gamma = 0.1$, $V = 0.05$, together with expected distributions for different level statistics.
			Level spacings are calculated per symmetry sector, in the middle $1/3$ of energies, but values from all sectors are included.
			\textbf{(b)}
			Finite size scaling of the mean value $\langle \tilde{r} \rangle$, for two different choices of $\Gamma$ and $V$.
			The dashed horizontal lines indicate the values corresponding to Poisson, semi-Poisson, and GOE level statistics.
		}
		\label{fig:level_statistics}
	\end{figure}

	In Fig.~\ref{fig:level_statistics}(a), the probability distribution $P(\tilde{r})$ of the ZXZ model is shown for the weakly interacting regime $\Gamma = 0.1$, $V = 0.05$.
	Data is obtained from the central third of the spectrum of ${\cal H}$ with system size $L = 18$.
	We find that $P(\tilde{r})$ exhibits clear level repulsion and is inconsistent with a Poissonian distribution.
	Thus, the small but nonzero $V$ is sufficient to break the integrability of ${\cal H}$.
	However, comparing $P(\tilde{r})$ to the theoretically expected distribution of a GOE, we also observe distinct deviations.
	Instead, it appears that $P(\tilde{r})$ is much better described by a so-called semi-Poisson distribution \cite{Garc_a_Garc_a_2006}, indicating that while ${\cal H}$ is nonintegrable, full quantum chaotic behavior is absent.
	This might be consistent with our findings from Figs.~\ref{fig:entropy} and \ref{fig:cluster_expectation}, i.e., that there are many states throughout the spectrum of ${\cal H}$ which appear to be not entirely in accord with the predictions of the ETH.
	In this context, let us note
	that it is well-known from other quantum scar models that
	$P(\tilde{r})$
	can in fact drift from semi-Poisson to GOE for increasing $L$
	\cite{Turner2018a}.
	
	To analyze the onset of quantum chaos further, Fig.~\ref{fig:level_statistics}(b) shows the mean value $\langle \tilde{r} \rangle$ versus system size $L$.
	In agreement with our previous observation in Fig.~\ref{fig:level_statistics}(a), we find that $\langle \tilde{r} \rangle$ does not reach the expected value $\langle \tilde{r} \rangle_\text{GOE} \approx 0.53$, instead seeming to approach the value for semi-Poisson statistics, although we can not rule out that the GOE value is eventually reached for even larger values of $L$.
	In contrast, if we consider slightly larger values of the perturbations, $V = 0.1$ and $\Gamma = 0.2$, we find a convincing agreement $\langle \tilde{r} \rangle \approx \langle \tilde{r} \rangle_\text{GOE}$ for $L = 16$.
	This suggests that full quantum chaos is restored in this parameter regime.
	Importantly, however,
	let us emphasize that the
	slow and atypical dynamics of some initial states occurs
	for these larger values of $\Gamma$ and $V$ as well, see Fig.~\ref{fig:Fig8}(b).
	\section{Details on low-lying
		excitations}\label{app::excitations}
	Let us define the following operators for periodic boundary conditions,
	\begin{align}
		\tilde{X}_\ell & = \left(K^+_{\ell} + K^-_{\ell}\right) = Z_\ell\ , \\ \tilde{Z}_\ell & = K_\ell\ .
	\end{align}
	These operators obey a spin algebra acting on the cluster basis of Appendix~\ref{app::cluster_basis}: $\tilde{X}_\ell$ inverts the cluster eigenvalue at site $\ell$, and $\tilde{Z}_{\ell}$ measures it.
	The Hamiltonian \eqref{eq:Ham} for $V = 0$ may be expressed purely in terms of these operators as,
	\begin{equation}
		{\cal H} = \hspace{-0.2cm}\sum_{\ell = 1}^{L/2-1} \left(\lambda \tilde{Z}_{2\ell} + \tilde{Z}_{2\ell+1}\right) + \Gamma \sum_{\ell = 1}^L \tilde{X}_{\ell - 1} \tilde{Z}_\ell \tilde{X}_{\ell + 1}\ .
	\end{equation}
	If we are in a regime such that $\Gamma \ll 1, \lambda$, then we can work in the space of a fixed number of cluster excitations.
	The action of the $\Gamma$ term is then to cause an excitation to hop to its next-nearest neighbor -- keeping to either the even or odd sublattice.
	Assume that there is just one excitation and that it is on the even sublattice.
	The effective Hamiltonian in this space is then,
	\begin{equation}
		\label{eq:effective_ham} {\cal H}_\text{eff} \ket{2\ell + 1} = 2\ket{2\ell + 1} - \Gamma\left(\ket{2\ell - 1)} + \ket{2\ell + 3}\right)\ ,
	\end{equation}
	where $\ket{\ell} = K^+_\ell \ket{\text{gs}}$ is a single excitation localized at site $\ell$, and energies are relative to the ground state.
	This can be solved by introducing momentum states $\ket{k}$ via a Fourier transform \eqref{eq:kstate},	where $k$ can take the values ${2\pi \eta}/({L/2})$ for integer $\eta$, $0 \leq \eta < L/2$.
	Applying this transform to Eq.~\eqref{eq:effective_ham}, and noting that the states $\ket{k}$ are mutually orthogonal, we obtain:
	\begin{equation}
		{\cal H}_\text{eff} \ket{k} = \left(2 - 2\Gamma\cos k\right) \ket{k}\ .
	\end{equation}
	This is the same result as for an excitation in the transverse-field Ising model, showing again the mapping between the ZXZ model and two copies of the TFIM.
	Note that with minor changes, similar excitations can be shown to exist as holes in a fully-excited sublattice, and also when the other sublattice is fully excited.
	\section{Nonzero momentum excitations}\label{App::kdata}
	\begin{figure}[tb]
		\centering
		\includegraphics[width=1\linewidth]{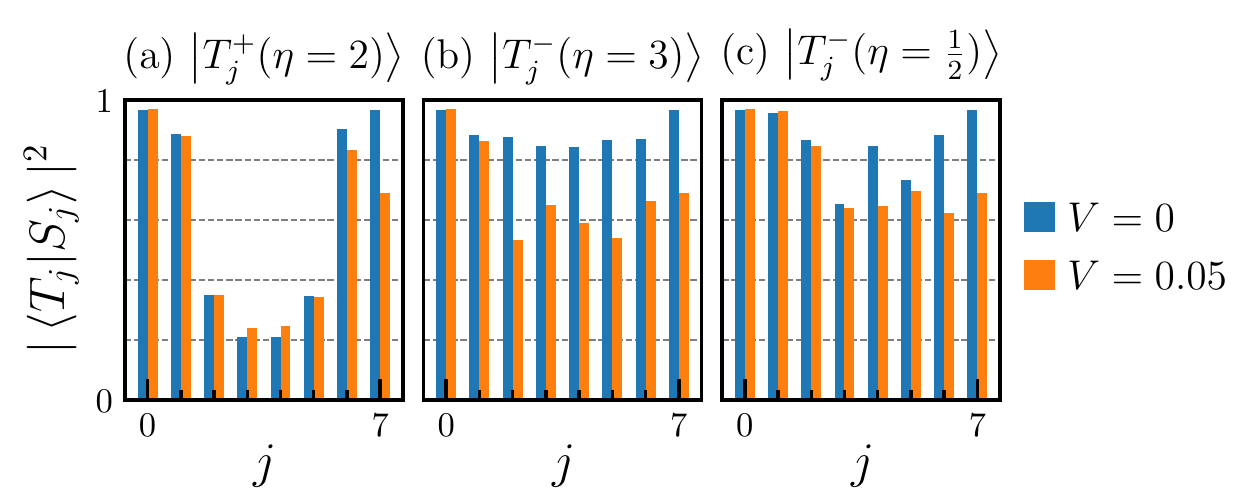}
		\caption{
			Overlaps of the states $\ket{S_j^\pm (k)}$ and 
			$\ket{T_j^\pm (k)}$, in analogy with the inset of Fig.~\ref{fig:tower}, for 
			states constructed with a generalized excitation operator $\mathcal{O}(k)$ 
			\eqref{eq:generalised_tower_operator}.
			Data for the non-interacting ($V=0$) model is in blue, 
			and for the interacting model ($V = 0.05$) in orange.
			[Other parameters: $L = 16$, $\lambda = 0.6$,
			$\Gamma = 0.1$, $G_\text{e} =
			G_\text{e} = +1$]
		}
		\label{fig:higher_momentum}
	\end{figure}
	We can generalize Eq.~\eqref{eq:tower_operator} to create excitations with an arbitrary momentum $k$ simply by introducing a location-dependent phase,
	\begin{equation}
		\label{eq:generalised_tower_operator} \mathcal{O}(k) = 
		\! \sum_{\ell = 1}^{L/2
			- 1}\! \!
		e^{ik\ell}
		K^+_{2\ell + 1}\ .
	\end{equation}
	
	Because the Hamiltonian is real symmetric, the eigenstates can be taken to be real.
	However, the towers $\ket{T_j (k)}$ formed from these operators are in general complex, so we must take linear combinations to produce real states.
	Since $\ket{T_j (k)}$ and $\ket{T_j (-k)}$ are complex conjugates, we can consider the states,
	\begin{equation}
		\ket{T_j^\pm (k)} = \frac{1}{\sqrt{2}} \left(\ket{T_j (k)} \pm \ket{T_j (-k)}\right)\ ,
	\end{equation}
	which are either real or have a global phase which can be eliminated by multiplication by a constant.
	We can also define $\ket{S_j^\pm (k)}$ in analogy with the definition in the main text.
	Note that $\ket{T_0 (k)} = \ket{\text{gs}}$, and $\ket{T_{L/2-1} (k)}$ is the state with a fully-excited odd sublattice and ground-state even sublattice, for all values of $k$, and so for these states we do not take linear combinations.
	
	Figure~\ref{fig:higher_momentum} shows exemplary data on the squared overlap, $\left|\braket{T_j^\pm (k)}{S_j^\pm (k)}\right|^2$, between these states and the eigenstates of the model.
	In Fig.~\ref{fig:higher_momentum}(a) and (b) we
	show data for $\eta = 2, 3$.
	For $\eta = 2$, the first excited state ($j=1$) and the penultimate state ($j = L/2 -1$) are close approximations, regardless of interactions, however the states in the middle of the tower are significantly worse.
	For $\eta = 3$, all the states are close approximations to eigenstates in the non-interacting model, and remain the dominant spectral contribution when interactions are turned on.
	The overlaps for other integer $\eta$ generally follow a similar pattern to one of these two cases.
	
	Figure~\ref{fig:higher_momentum}(c) shows that
	half-integer values of $\eta$ can produce good overlaps too;
	in particular, we
	find that for $\eta = 1/2$ the overlap for the first
	excited
	state is very high, and even better than for $k = 0$ in the main text.
	
	\section{Matrix Product State (MPS) representation of the tower of states}\label{App:MPS}
\begin{figure}[tb]
		\centering
		\includegraphics[width=0.9\columnwidth]{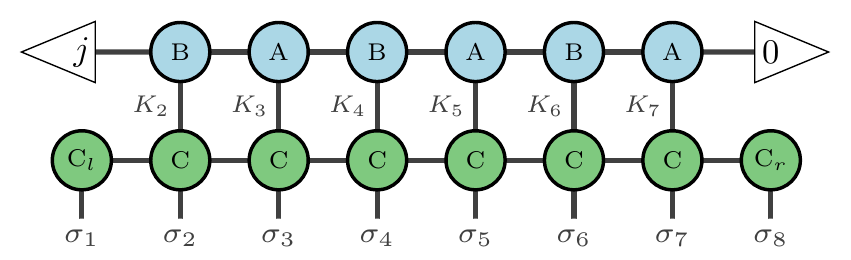}
		\caption{
			Matrix product state representation of a tower state $\ket{T_j (k)}$.
			The tensors $\mathrm{A}, \mathrm{B}$ generate $j$ delocalised excitations with momentum $k$ on the odd sublattice, and the tensors $\mathrm{C}$ map this from a cluster basis description to the physical spin basis.
			When the vertical edges $K_\ell$ are contracted, we obtain an MPS of maximal bond dimension $\chi = 4\left[\max(j + 1, L/2 - j)\right] \leq L + 4$.
			Triangles indicate legs which are held constant (achieved by contracting with boundary vectors $A_l$ and $A_r$). 
		}
		\label{fig:tensor_network}
	\end{figure}

	In this section we show that the states $\ket{T_j (k)}$ in fact admit an MPS representation with a maximum bond dimension linear in $L$.
	We do this in two steps; see Fig.~\ref{fig:tensor_network} for reference.
	First, we construct $j$ cluster-wave excitations of momentum $k$ on the odd sublattice (tensors $A, B$, in blue), and then we map that state from the cluster basis to the spin basis via a matrix-product operator (MPO) of dimension $\chi=4$ (tensors $C$, in green).
	
	We demonstrate this construction first for the simple case $j=1$, i.e.\ a single delocalized cluster excitation on the odd sublattice.
	We can write this state in the cluster basis (App.~\ref{app::cluster_basis}) as,
	\begin{equation}
		\ket{\psi} = \sum_{\ell=1}^{L/2-1} e^{ik\ell}
		\ket{\bar{1}\bar{1}\cdots1_{2\ell + 1} \cdots\bar{1}\bar{1}}\ .
	\end{equation}
	We can then write this in matrix product form as,
	\begin{gather}
		\label{eq:MPS_def}
		\ket{\psi} = \sum_{\left\{K_\ell\right\}} A_l\!
		\left(\prod_{\ell=1}^{L/2 - 1} 
		\mathrm{B}_{K_{2\ell}} \!\! \mathrm{A}_{K_{2\ell + 1}}
		\right)\! A_r
		\ket{K_2 K_3 \cdots K_{L - 1}}\ ,\\
		\label{eq:MPS_simple} 
		\mathrm{A}_{+1} = \begin{pmatrix} 0 & 0\\1 & 0 \end{pmatrix}\ ,\ 
		\mathrm{A}_{-1} = \begin{pmatrix} 1 & 0\\0 & \alpha \end{pmatrix}\ ,
	\end{gather}
	where $A_l = (0, 1)$ and $A_r = (1, 0)^\mathrm{T}$, $\alpha = e^{ik}$, and $\mathrm{B}_K = \mathbb{I} \delta_{K, -1}$ selects states with $K_{2\ell} = -1$ on even sites.
	
	Consider a single element of the summation in Eq.~\eqref{eq:MPS_def}.
	Note that $\mathrm{A}_{-1}^n A_r = A_r$ but $\mathrm{A}_{-1}^n \mathrm{A}_{+1} A_r = \alpha^n A_l^\mathrm{T}$, while $\mathrm{A}_{+1}^2 = 0$.
	Hence it is clear there must be exactly one $\mathrm{A}_{+1}$ in the matrix product or else it will vanish, while we accumulate a factor $\alpha$ for each site to the left of this $\mathrm{A}_{+1}$.
	This gives the desired state, up to a constant phase factor which we can ignore.
	
	In the above example, the index $n$ of the nonzero entry in the vector as we moved from right to left counted the number of excitations to the right of the current site.
	To extend this idea to $j$ such excitations, we need a bond dimension of $j + 1$ such that $0 \leq n \leq j$,
	\begin{equation}
		\mathrm{A}_{+1} = \scalebox{0.7}{$
			\begin{pmatrix}
				0 &&&& \\
				1 & 0 &&& \\
				& \alpha & \ddots && \\
				&& \ddots & 0 & \\
				&&& \alpha^{j-1} & 0 \\
			\end{pmatrix}
			$}\ ,\
		\mathrm{A}_{-1} = \scalebox{0.8}{$
			\begin{pmatrix}
				1 &&&& \\
				& \alpha &&& \\
				&& \alpha^2 && \\
				&&& \ddots & \\
				&&&& \alpha^j\\
			\end{pmatrix}
			$}\ ,
	\end{equation}
	where we now take $A_l = (1, 0, \dots, 0)$ and $A_r = (0, \dots, 0, 1)^\mathrm{T}$.
	Note that now a factor $\alpha^n$ is accumulated at each site when there are $n$ excitations to the right. 
	
	We now transform this state $\ket{\psi} = \ket{T_j (k)}$ to the physical spin basis using an MPO formed of the rank-4 tensors $\mathrm{C}^K_\sigma = \operatorname{diag}\!\left(\mathrm{c}^K_\sigma, \mathrm{c}^K_\sigma\right)$ where,
	\begin{gather}
		\mathrm{c}^K_\downarrow =
		\begin{pmatrix}
			0 & K \\ 0 & -K
		\end{pmatrix}
		\ ,\ 
		\mathrm{c}^K_\uparrow =
		\begin{pmatrix}
			1 & 0 \\ 1 & 0
		\end{pmatrix}
		\ .%\ 
	\end{gather}%
	The tensors $\mathrm{c}^K_\updownarrow$ here form the known MPS description for a cluster state, where $K = \pm 1$ is the eigenvalue of $K_\ell$ for the state at each site~\cite{Perez-Garcia2006a}.
	Finally, the tensors $\mathrm{C}_l$ and $\mathrm{C}_r$ are the contraction of the left or right leg respectively of $\mathrm{C}^{+1}_\sigma$ with an appropriate boundary vector -- this determines the symmetry sector.
	This MPO hence has bond dimension $\chi_C = 4$ 
	
	When the vertical edges representing $K_\ell$ are contracted, the resultant MPS will have maximal bond dimension $\chi = \chi_A \chi_C = 4(j+1)$, which is $O(L)$ since $j < L/2$.
	Certain optimizations can improve this to $4\max(j+1, L/2 - j)$.
	This bounds the entropy growth of the $|T_j(k)\rangle$ by $O(\log L)$.
	
	\section{Spectral decomposition of cluster-basis states}
	\label{App::Overlap}
	\begin{figure}[tb]
		\centering
		\includegraphics[width=1\columnwidth]{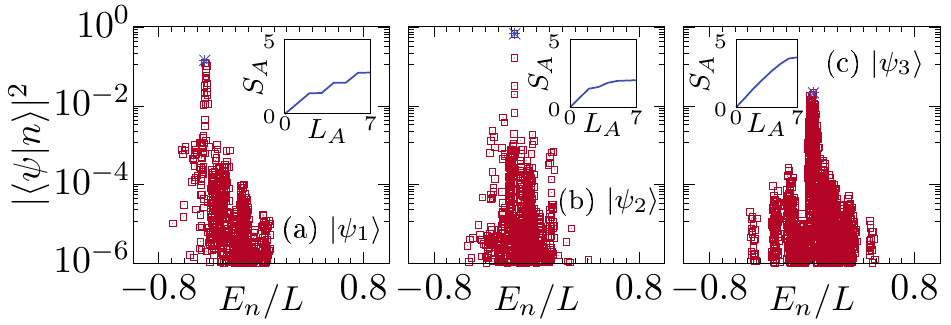}
		\caption{
			Overlap of the states $\ket{\psi_1}$-$\ket{\psi_3}$ (cf.\ Fig.~\ref{FigProfile}) with the eigenstates $\ket{n}$ of ${\cal H}$.
			The insets show the entanglement entropy $S_{A}$ versus subsystem size for the eigenstate $\ket{n}$ which has the largest overlap (indicated by the blue asterisk).
			The parameters are chosen as $\Gamma = 0.1$, $V = 0.05$, and $\lambda = 0.6$.
			The system size is $L = 14$.
		}
		\label{FigOverlap}
	\end{figure}
	In Figs.~\ref{FigOverlap}(a)-(c), we plot the overlap $|\langle \psi|n\rangle|^2$ of the three cluster-basis states $\ket{\psi_1}$-$\ket{\psi_3}$ (cf.
	\ Fig.~\ref{FigProfile}) with the
	eigenstates $\ket{n}$ of ${\cal H}$.
	While the spectral weight of each $\ket{\psi}$ is different, we here particularly focus on the eigenstate $\ket{n}$ which shows the largest overlap with the respective $\ket{\psi}$ (marked by an asterisk).
	For this particular $\ket{n}$, the insets in Figs.~\ref{FigOverlap}(a)-(c) show its entanglement entropy $S_{A}$ versus subsystem size.
	While an area-law (or at least sub-volume-law) is observed for the first two states, we find a clear volume law of $\ket{n}$ in the case of the quickly thermalizing state $\ket{\psi_3}$.
	The atypical dynamics of some of the initial states thus appears to be linked to the existence of and their overlap with low-entangled eigenstates in the spectrum of ${\cal H}$.
	
	\bibliography{ZXZ}
\end{document}